\documentclass[lettersize,journal]{IEEEtran}
\IEEEoverridecommandlockouts
\usepackage{cite}
\usepackage{indentfirst}
\usepackage{graphicx}
\usepackage{svg}
\usepackage{changepage}
\usepackage{stfloats}
\usepackage{amsfonts,amssymb}
\usepackage{bm}
\usepackage{algorithm}
\usepackage{algpseudocode}
\usepackage{amsmath}
\usepackage{amsmath,amssymb,amsfonts}
\usepackage{times}
\usepackage{url}
\usepackage{cite}
\usepackage{textcomp}
\usepackage{xcolor}
\usepackage{color} 
\usepackage{setspace}
\usepackage{enumitem}
\usepackage{float}
\usepackage{diagbox}
\usepackage[T1]{fontenc}
\usepackage[utf8]{inputenc}
\usepackage{threeparttable}
\usepackage{booktabs}
\usepackage{footnote}
\usepackage{graphicx}
\usepackage{pifont}
\usepackage{subfigure}
\usepackage{mathrsfs}
\usepackage{makecell}
\usepackage{amsmath}
\usepackage[justification=centering]{caption}
\usepackage{lipsum}
\usepackage[numbers,sort&compress]{natbib}
\usepackage{amsmath}

\newtheorem{remark}{\textbf{Remark}}
\newtheorem{definition}{\textbf{Definition}}

\newcommand{\adots}
\newenvironment{proof}{{\indent  \indent \it Proof:}}{\hfill $\blacksquare$}

\begin{document}
\title{Multi-scale Vehicle Localization In Heterogeneous Mobile Communication Networks}

\author{Lele Cong,
	Kaitao~Meng,~\IEEEmembership{Member,~IEEE},
	Deshi~Li,	
	Hao~Jiang,~\IEEEmembership{Member,~IEEE},
	and Liang~Xu 
	% <-this % stops a space
	\thanks{Lele Cong, Deshi Li, Hao Jiang, and Liang Xu are with the School of Electronic Information, Wuhan University, Wuhan 430072, China. (emails: conglele@whu.edu.cn; dsli@whu.edu.cn; jh@whu.edu.cn; lgxu@whu.edu.cn).
		
		Kaitao Meng is with the Department of Electronic and Electrical Engineering, University College London, UK (email: kaitao.meng@ucl.ac.uk).
	}
}

%\textit{(Corresponding author: .)}

\maketitle

\vspace{-10mm}
\begin{abstract}
%核心创新点，采用三个定位粒度降低定位空间维度，同时每个粒度内的位置间隔是不同的，这是为了提取最能代表每个位置区域的信号特征，通过准确的特征匹配实现多维度的定位。
%交代背景，现有研究的不足
Low-latency and high-precision vehicle localization plays a significant role in enhancing traffic safety and improving traffic management for intelligent transportation. However, in complex road environments, the low latency and high precision requirements could not always be fulfilled due to the high complexity of localization computation. To tackle this issue, we propose a road-aware localization mechanism in heterogeneous networks (HetNet) of the mobile communication system, which enables real-time acquisition of vehicular position information, including the vehicular current road, segment within the road, and coordinates. By employing this multi-scale localization approach, the computational complexity can be greatly reduced while ensuring accurate positioning. Specifically, to reduce positioning search complexity and ensure positioning precision, roads are partitioned into low-dimensional segments with unequal lengths by the proposed singular point (SP) segmentation method. To reduce feature-matching complexity, distinctive salient features (SFs) are extracted sparsely representing roads and segments, which can eliminate redundant features while maximizing the feature information gain. The Cramér-Rao Lower Bound (CRLB) of vehicle positioning errors is derived to verify the positioning accuracy improvement brought from the segment partition and SF extraction. Additionally, through SF matching by integrating the inclusion and adjacency position relationships, a multi-scale vehicle localization (MSVL) algorithm is proposed to identify vehicular road signal patterns and determine the real-time segment and coordinates. Simulation results show that the proposed multi-scale localization mechanism can achieve lower latency and high precision compared to the benchmark schemes.
\end{abstract}

\begin{IEEEkeywords}
vehicle localization, low latency, multiple scales, salient feature (SF), heterogeneous networks (HetNet)
\end{IEEEkeywords}

\section{Introduction}
%为什么要研究道路定位（为什么道路定位重要，有什么实际价值），GPS和INS解决道路定位存在的问题
\par
Vehicle localization plays a vital role to enhance road safety, optimize path planning, and improve traffic management for intelligent transportation \cite{Yu2022Precise}. There is a high demand for vehicle localization services on roads, where vehicle localization accounted for more than 90$\%$ of road-related localization services \cite{requirement}. Especially, to ensure road traffic safety, low-latency and high-precision localization is essential for high-speed vehicles. However, low-latency vehicular localization could not always be fulfilled without sacrificing positioning precision due to complex positioning computation. With the integration of the global positioning system (GPS) inside vehicles, satellite navigation technology has been widely used for vehicle localization. However, the responding time and positioning precision of GPS are affected by satellite transmission path loss, especially in severe weather and dense obstructions scenarios \cite{Vo2016Survey}. Although the inertial navigation system (INS) can locate vehicles through dead reckoning, the long-term state prediction would inevitably lead to cumulative errors, which may decrease localization precision \cite{Jian2018MEMS}. Considering that the position of the in-vehicle communication equipment and the vehicular user equipment (VUE) remains relatively stationary with the vehicular position, the vehicular position can be equivalently estimated by locating the VUE. With seamless coverage and accessibility, cellular networks can provide alternative opportunities for VUE localization to reduce latency and guarantee localization precision \cite{2018Survey}.
\par
Based on cellular network signals, three typical positioning methods can be used for VUE localization, i.e., Cell ID-based methods \cite{Zakaria2017PerformanceEO, Floarea2020COO, Li2022E-CID, Bore2005ECA,Lao2018survey}, geometric-based methods \cite{2021overview,2023Kaitao,Abbas2018NLOS, Gen2020IDEN, Huang2018IDEN,Carpi2019IDEN, Xiao2017INEN,Van2015IDEN,Mara2010NLOS,Yanlong2011TOA,Chan1994est,Xiong2015IDEN}, and fingerprint-based methods \cite{Giovanni2018Passive, 2018IOT,Yang2019TDOA, 2021IOT,Tian2020Subspace,2020K-MeansFingerprint, Mo2012Clustering,9380161}. Though the delay of Cell ID-based methods can be shortened by obtaining a serving cell position, the accuracy relies on the serving cell radius \cite{Zakaria2017PerformanceEO}. To enhance the localization precision, radio frequency (RF) signal information, such as received signal strength (RSS), time of arrival (TOA), time difference of arrival (TDOA), or angle of arrival (AOA) can be utilized for VUE positioning through geometric-based methods \cite{2021overview}. Nevertheless, since the explicit line-of-sight (LoS) links between base stations (BSs) and VUE are difficult to maintain in complex road environments, e.g., tunnels and overpasses, the localization accuracy of geometric-based methods may be severely degraded \cite{2023Kaitao}. Moreover, multiple distance estimation computations between VUE and BSs result in intolerable response delays \cite{Abbas2018NLOS}. 
%给出信号匹配定位的好处
Without relying on complex signal information and LoS links, fine-grained RF signals from multiple transmitters, e.g., Wi-Fi access points (APs) and BSs can be extracted to locate VUE through fingerprint matching \cite{Giovanni2018Passive}, \cite{2018IOT}. However, fine-grained fingerprint matching is extremely time-consuming to ensure vehicular localization precision in wide outdoor environments \cite{Yang2019TDOA}, \cite{2021IOT}. Additionally, the faint decay of millimeter waves according to various environments would greatly affect localization performance \cite{2017HetNet}. With rapidly growing demands for high-reliability and low-latency mobile communication services, the heterogeneous networks (HetNet) BSs including the fourth generation (4G) and the fifth generation (5G) BSs have been widely deployed to improve network performance \cite{2021HetNet}. Recent research in mobile networks tended to strengthen localization precision by intelligent reflective surfaces \cite{10226306} and ultra-massive MIMO \cite{Hao2024}, etc., while overlooking the potential of HetNet to reduce localization latency \cite{10199909}. Moreover, there are different fluctuation patterns of HetNet signals since the signal diffraction capabilities and path loss vary with frequency bands, which presents valuable opportunities for effective vehicle positioning. Therefore, studying vehicle localization in HetNet is urgently needed.
\par
%挑战，道路定位困难，1基于单个基站信号，由于传播距离传播环境变化，从整个道路中提取无处不在的特征困难 2对于空间交叠与平行的道路，道路信号相似，仅依靠道路尺度的信号特征很难准确定位道路位置
%解决办法，1基于异构基站定位，2采用多尺度信号特征提取，不依赖视距环境
Due to relatively constant positions of signal blockage obstacles along roads, e.g., trees and buildings, signals between BSs and VUE orderly switch between LoS and non-line-of-sight (NLoS) links \cite{10124080}. Hence, distinctive statistical signal features of HetNet RSS values, e.g., the mean value, the variance value, the signal gradient, etc., can be extracted to represent corresponding road sections. Then, signal feature matching is designed for segment positioning by minimizing the Euclidean distance between the offline extracted signal features and the real-time signal features. However, achieving both low latency and high precision is still challenging due to the following reasons. First, achieving high precision requires fine-grained feature matching, leading to higher computational complexity and longer localization response time. Second, coarse position granularity with sparse signal features could provide low positioning latency, but resulting in high localization errors. Hence, it is critical to reduce feature-matching complexity and ensure localization precision at the same time. 
%key idea
Due to the varying spatial distances, carrier frequencies, and signal propagation environments between VUE and HetNet BSs, significant differences in signal gradients can be observed across different road sections, as illustrated in Fig. \ref{figure1}, where a \textbf{\textit{segment}} can be defined as a road section obtained by partitioning the road-scale HetNet signal sequence based on signal gradients. Therefore, high-dimensional coordinates within a road can be grouped into low-dimensional segments with different signal gradients, which can reduce the complexity of position searching within roads and improve road localization precision. As not all statistical signal features significantly differ between roads and segments. To sparsely and accurately represent roads on various scales, \textbf{\textit{salient features (SFs)}} are defined as the statistical features with significant differences among roads and segments, where the significant difference refers to the absolute difference of the normalized features between two adjacent positions exceeding a predetermined threshold. To the best of our knowledge, this is the first time that low-dimensional salient features extraction and multi-scale road representation are realized to locate vehicles in complex urban road environments, which can reduce localization computational complexity and ensure accuracy at the same time. The main contributions in this paper can be summarized as follows:
\begin{itemize}
	\item We propose a road-aware localization mechanism in HetNet, which includes the HetNet signal feature extraction and multi-scale vehicle localization to enable real-time acquisition of vehicular position information on different spatial scales, i.e., roads, segments, and coordinates. 
	\item We propose a singular point (SP) segmentation method extracting gradient change points for road segmentation, based on which we then propose a salient feature (SF) extraction method by maximizing feature information gain to extract signal features with significant differences. Moreover, to theoretically verify the improvement of localization accuracy by road partitioning and SF extracting, the Cramér-Rao Lower Bound (CRLB) is derived.  
	\item We propose a multi-scale vehicle localization (MSVL) algorithm, which includes three modules, i.e., the road recognition module, the segment positioning module, and the coordinate localization module. By integrating the spatial inclusion and adjacency properties of roads and segments, the MSVL algorithm can quickly determine the vehicle's multi-scale positions with high precision based on SF matching. 	  
	\item We conduct real-world experiments to verify the proposed multi-scale localization mechanism in a real HetNet scenario. The experimental results reveal that our proposed mechanism can achieve high positioning precision with low delay.
\end{itemize}
%%%%%%%%%%%%%%%%%%%%%%%%%%%%%%%%%%%%%%%%%%%%%%%%%%%%%%%%%%%%%%%%%%%%%%%%%%%%%%%
\par
The remainder of this paper is organized as follows. Section \ref{Relate Work} summarizes the related work and Section \ref{SYSTEM} describes the proposed mechanism model and problem formulation. In section \ref{Offline}, HetNet signal sequences are segmented to construct multi-scale positioning space, and SFs are extracted to sparsely represent roads and segments. The multi-scale localization algorithm is designed in section \ref{Online}. Section \ref{Experiment Results} provides numerical results to reveal the performance of our proposed mechanism. Section \ref{Conclusions} concludes this paper.
\par
\textit{Notations}: A capital bold-face letter denotes a matrix, e.g., $\boldsymbol{W}$. A vector is denoted by a bold-face lowercase letter, e.g., $\boldsymbol{f}$. The $\mathit{l}_2$ norm of a vector is represented by $\lVert\cdot\rVert_2$. The absolute value of a variable is denoted as $\left| \cdot \right|$.
\section{Relate Work}
\label{Relate Work}
This section provides an overview of the current literature on cellular network-based localization, specifically focusing on the performance in terms of latency and precision. Three kinds of typical positioning methods based on cellular networks have been investigated, i.e., Cell ID-based methods \cite{Floarea2020COO,Li2022E-CID, Bore2005ECA,Lao2018survey}, geometric-based methods \cite{Gen2020IDEN, Huang2018IDEN,Carpi2019IDEN, Xiao2017INEN,Van2015IDEN,Mara2010NLOS,Yanlong2011TOA,Chan1994est,Xiong2015IDEN}, and  fingerprint-based methods \cite{Tian2020Subspace,2020K-MeansFingerprint, Mo2012Clustering,9380161}. 
\subsection{Cell ID-based Method}
Cell ID-based methods localize VUE by obtaining the ID of the VUE's service cell \cite{Floarea2020COO}. Though high positioning speed can be achieved due to low computational complexity, the localization precision depends on radius of BS coverage areas. To reduce localization errors, supplementary data such as timing advance (TA) and round-trip time (RTT) are utilized in enhanced Cell-ID methods \cite{Li2022E-CID}, \cite{ Bore2005ECA}. Despite this, the enhanced Cell-ID methods remain relatively coarse in precision, typically within tens of meters in urban small-cell networks and several hundred meters in rural macro-cell networks \cite{Lao2018survey}.   
\begin{figure}[t]
	\centering
	\includegraphics[width=8.5cm]{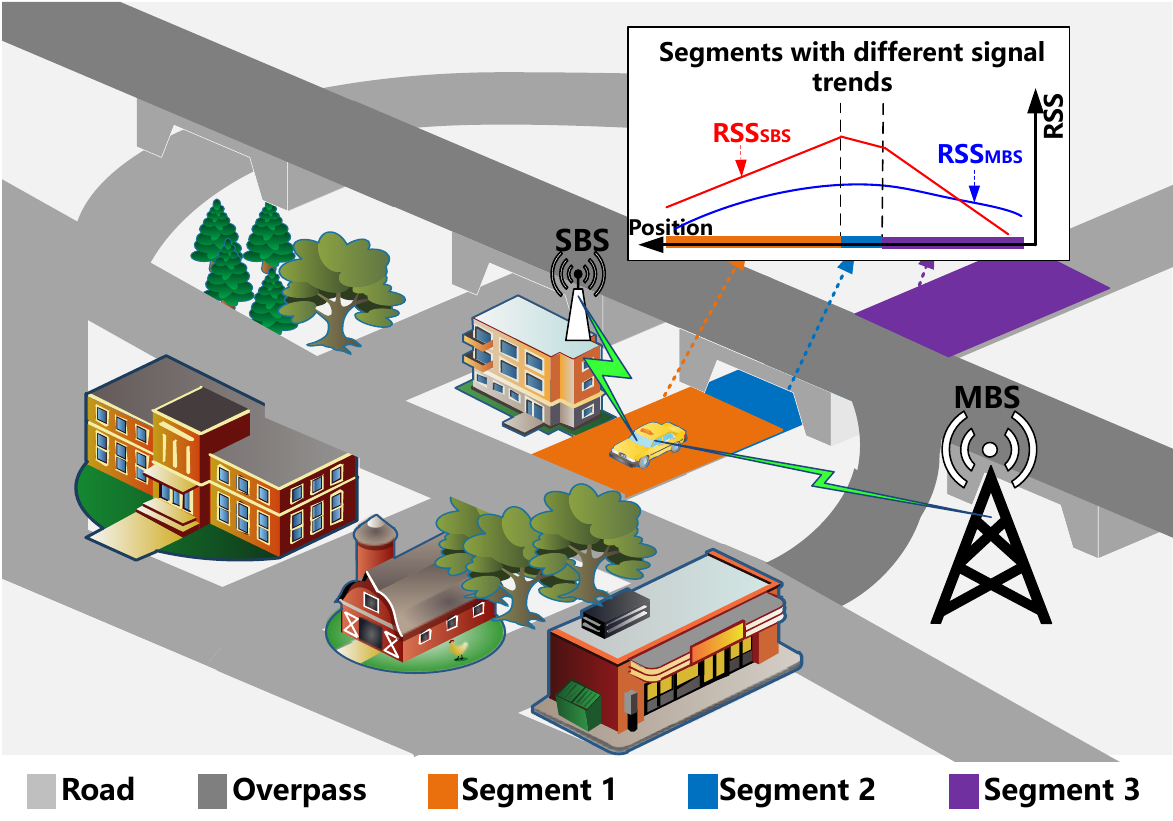}
	\vspace{-1mm}
	\caption{Road localization scenario in urban area.}
	\vspace{-3mm}
	\label{figure1}
\end{figure}
\subsection{Geometric-based Method}
To improve localization accuracy, geometric-based methods can be used to localize vehicles by distance or angle estimating between VUE and multiple BSs. Nevertheless, the precision of geometric-based methods is compromised by NLoS links of any BS signals. To reduce the localization errors, the NLoS measurements can be first distinguished by machine learning approaches, e.g., support vector machine (SVM) \cite{Gen2020IDEN, Huang2018IDEN}, neural network (NN) \cite{Carpi2019IDEN, Xiao2017INEN}, and relevance vector machine (RVM) \cite{Van2015IDEN}. Then, the localization errors of NLoS links can be mitigated using the least squares method \cite{Mara2010NLOS}, the constrained optimization method \cite{Yanlong2011TOA}, the maximum likelihood method \cite{Chan1994est}, and the robust statistics method \cite{Xiong2015IDEN}. However, reducing positioning errors incurs a significant time overhead that increases exponentially as the number of BSs involved grows \cite{Abbas2018NLOS}. Additionally, suppose that LoS links exist between all BSs and the vehicle, intolerable positioning delays may result from multiple distance estimations between the VUE and all BSs. 
\vspace{-1mm}
\subsection{Fingerprint-based Method}
Independent of LoS environments, a fingerprint is defined as a vector of signal strengths from multiple transmitters at grids of known locations, i.e., reference points (RPs) \cite{8931646}. Then, the basic idea of a fingerprinting approach is to locate a mobile device by comparing its RSSs to a predefined database of fingerprints \cite{8392372}. Fingerprint-based localization is effective for indoor environments, where GPS signals may be absent or unreliable due to signal blockage or multi-path effects \cite{2018Survey}. However, due to fine-grained fingerprint matching, the high computation complexity results in intolerable localization delays. To reduce positioning latency and online computation costs, clustering methods have been proposed to group fingerprints \cite{2020K-MeansFingerprint}, \cite{Mo2012Clustering}. The author of paper \cite{Tian2020Subspace} utilized subspace-based techniques to compress the dimension of its fingerprint database. However, on the one hand, a huge amount of human labor would be consumed to collect fingerprints in wide outdoor road environments. On the other hand, a large fingerprint database would consume huge maintenance and update costs if transmitters' numbers, positions, or device types changed \cite{9380161}. In addition, high fingerprinting localization precision depends on RF signals from as many transmitters as possible, which may not always be fulfilled due to the severe signal attenuation in the 5G and 6G mobile communication systems.  
\par
The existing positioning methods based on cellular networks mainly focus on reducing positioning errors by complex localization algorithms and ignore positioning delay performance. However, to ensure the traffic safety of high-speed vehicles, both low-latency and high-precision positioning performance are indispensable. Unlike previous positioning methods, we propose a road-aware positioning mechanism by establishing a multi-scale positioning space and extracting salient features to reduce positioning latency while ensuring positioning accuracy.    
\section{Mechanism Modeling and Problem Formulation}
\label{SYSTEM}
\subsection{Road-aware Localization Mechanism}
\begin{figure}[t]
	\centering
	\setlength{\abovecaptionskip}{0.cm}
	\includegraphics[width=8.5cm]{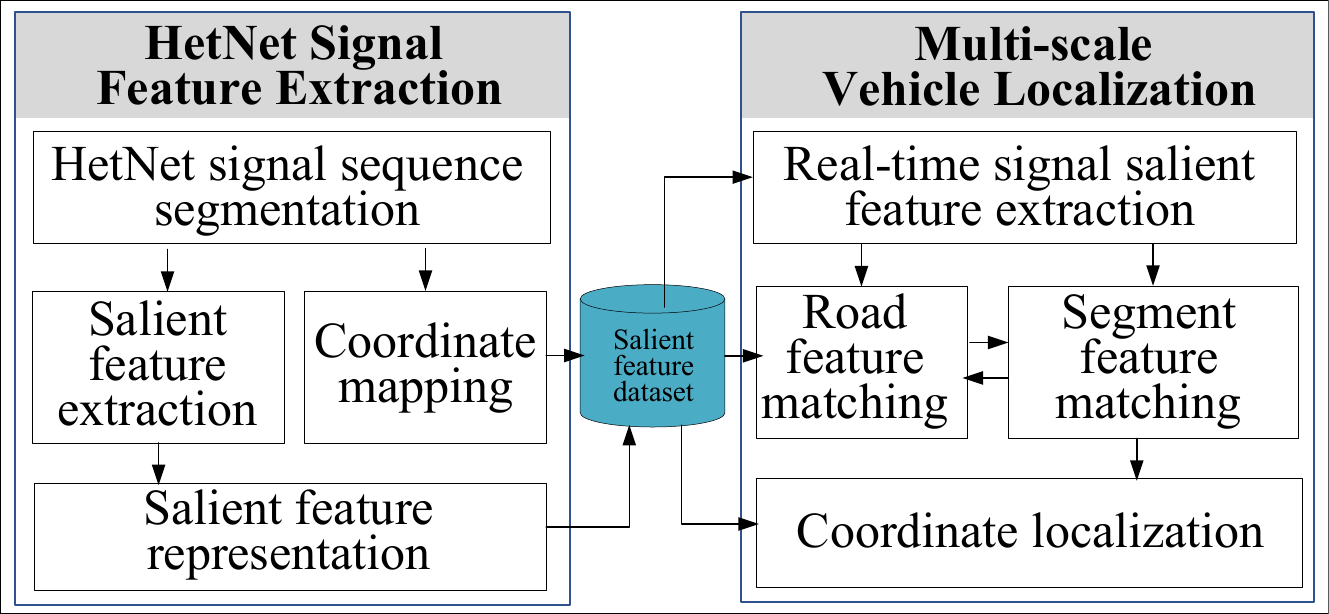}
	\vspace{1mm}
	\caption{Road-aware localization mechanism.}
	\vspace{-3mm}
	\label{figure2}
\end{figure}

In complex urban road environments portrayed in Fig. \ref{figure1}, to achieve low-latency localization, we propose a road-aware localization mechanism with the architecture illustrated in Fig. \ref{figure2}. By analyzing the fluctuation trends of HetNet signals under different NLoS and LoS signal propagation conditions, wide road environments can be divided into low-dimensional segments, which construct the multi-scale positioning space including roads and segments. Moreover, due to the change in signal propagation environments and signal propagation distances, distinctive signal features can be extracted for position mapping. The proposed mechanism includes two stages: the HetNet signal feature extraction and the multi-scale vehicle localization.
\par
\emph{HetNet signal feature extraction}: To construct the low-dimension positioning space on multiple scales, roads can be partitioned into segments through HetNet signal sequence segmentation, which would reduce the positioning computation complexity. By extracting SFs from segmented HetNet signal sequences, roads and segments can be sparsely represented to reduce feature-matching complexity. Moreover, the CRLB is derived to verify the positioning accuracy improvement brought by the proposed segment partition and SF extraction.   
\par
\emph{Multi-scale vehicle localization}: To accurately locate a vehicle's position on roads, the real-time SFs of the vehicular HetNet signals will be extracted. Then, based on road and segment SF matching, a MSVL algorithm is designed to quickly obtain vehicular real-time positions on roads. To improve vehicle positioning precision, the coordinates within a segment can be localized based on fitted curves.
%定位问题建模，场景（异构基站，道路场景），信道模型（功率，特征），问题模型（概率，匹配）
\subsection{HetNet Signal Representation}
%多空间尺度描述，信道模型描述，多空间尺度特征说明
To form road-scale HetNet signal sequences extracting signal features, roads within a SBS coverage area can be defined as a road positioning space $\mathcal{R}$, i.e., $\mathcal{R}=\left\{\rm{r}_1,\rm{r}_2,\hdots,\rm{r}_\textit{m} \right\}$, where $\rm{r}_\textit{i}$ is a road in $\mathcal{R}$, and the number of roads is represented as $m$. Within the road positioning space $\mathcal{R}$, a road-scale HetNet signal sequence is defined as a vector $\boldsymbol{\phi}_i$ which is denoted as $\boldsymbol{\phi}_i = \left[\boldsymbol{o}_1, \boldsymbol{o}_2, \hdots, \boldsymbol{o}_{L_i}\right]^\top$, where $\boldsymbol{o}_j$ represents a HetNet signal vector of the $j$th position on the road $\rm{r}_\textit{i}$. In $\boldsymbol{\phi}_i$, $L_i$ denotes the number of signal sampling positions within the road $\rm{r}_\textit{i}$.  $L_i$ is determined by the signal sampling distance interval $\Delta D$. To ensure consistent measurement granularity of signal features for effective road representation and identification, $\Delta D$ remains constant across different roads and segments. This signal sampling distance interval is influenced by the time interval of the signal sampling and the vehicular velocity. The HetNet signal vector $\boldsymbol{o}_j$ takes the form of $\boldsymbol{o}_j=\left[P_{j,1}, P_{j,2},\hdots, P_{j, K}\right]^\top$, where $P_{j,k}$ denotes the RSS of the $k$th BS. The number of HetNet BSs is defined as $K$ including 1 macro BS (MBS) and $K-1$ small-cell BSs (SBSs). Based on road-scale HetNet signal sequences and corresponding signal features, our proposed mechanism is designed independent of BS locations and transmit powers locating vehicles, which can save data storage space and computation costs. The coordinate vector of the $j$th position is denoted by $\boldsymbol{c}_j=\left(x_j,y_j\right)^\top$, where the $x_j$ and $y_j$ respectively represent 2 dimensional (2D) coordinates, i.e., longitude and latitude. The RSS of the $k$th BS $P_{j,k}$ (in dBm) on the $j$th position is given by \cite{optimal}:
\begin{equation}\label{plm}
\vspace{-2mm}
	P_{j,k} = P_0-10{\beta_k}\log_{10}\Vert \boldsymbol{c}_j-\boldsymbol{c}_{BS_k} \Vert_2+X_k,		
\vspace{1mm}
\end{equation}
where $P_0$ denotes the original RSS at the target, and $\beta_k$ represents the path loss parameter depending on signal propagation environments. $X_k$ is the Gaussian white noise (GWN). In (\ref{plm}), the position of the $k$th BS is denoted by the coordinate vector  $\boldsymbol{c}_{BS_k}=\left(x_{BS_k},y_{BS_k}\right)^\top$. The signal gradient ${g_{_{j,k}}}$ of the $k$th HetNet BS can be calculated by the RSS difference value per unit distance, which can be expressed as follows: 
\begin{equation}\label{g}
	%	\vspace{-1mm}
	\setlength\abovedisplayskip{3pt}
	g_{_{j,k}}=  \frac{P_{j+1,k}-P_{j,k}}{\left(\left(x_{j+1}-x_j\right)^2+\left(y_{j+1}-y_j\right)^2\right)^{\frac{1}{2}}}
	.
	\vspace{-1mm}
\end{equation} 
\indent
Within the road $\rm{r}_\textit{i}$, a segment positioning space $\mathcal{S}_i$ can be established by partitioning HetNet signal sequence based on signal gradients, i.e., $\mathcal{S}_i=\left\{ {\rm{s}_1},{\rm{s}_2},\hdots,{\rm{s}}_{n_i+1} \right\}$, where $\rm{s}_\textit{l}$ represents the $l$th segment corresponding to a segment-scale HetNet signal sequence $\boldsymbol{\psi}_l$, and $n_i+1$ represents the segment number. The $l$th segment can be represented by the 2D coordinate vector $\left(x_{m}^l,y_{m}^l\right)^\top$, where $x_{m}^l$ and $y_{m}^l$ respectively denote the longitude and latitude of the middle position within this segment. Within the segment $\rm{s}_\textit{l}$, a coordinate space is defined as $\mathcal{C}_l$ composed by all RSS sampling coordinates, i.e., $\mathcal{C}_l = \left\{\boldsymbol{c}_1,\boldsymbol{c}_2,\hdots,\boldsymbol{c}_{N_l} \right\}$, where $N_l$ represents the coordinate number of the segment ${\rm{s}}_l$. Then, a multi-scale positioning space is established, consisting of roads, segments, and coordinates. In this multi-scale positioning space, the vehicular position vector can be expressed as $\boldsymbol{p} = \left[{\rm{r}_\textit{i}},{\rm{s}_\textit{l}},\boldsymbol{c}_j\right]^\top$. Based on corresponding HetNet signal sequences, the road-scale and segment-scale SF vectors, i.e., $\boldsymbol{f}_{\rm{r}_\textit{i}}$ and $\boldsymbol{f}_{\rm{s}_\textit{l}}$, can be extracted to represent road ${\rm{r}_\textit{i}}$ and segment ${\rm{s}_\textit{l}}$, respectively. Most notations used in this paper are listed in Table \ref{tab:tab1}.
\begin{table}[h]	
	\centering
	\captionsetup{font=small}
%	\caption{IMPORTANT NOTIFICATIONS AND SYMBOLS USED IN THIS WORK}
	\caption{\sc \\
		Important Notifications and Symbols Used in This Work}\label{tab:tab1}
	\vspace{-2mm}
	\begin{tabular}{l  l}
		\hline
		\textbf{Notation} & \textbf{Physical meaning}\\\hline 
		$\rm{r}_\textit{i}$ & Road $i$  \\\hline 
		$\rm{s}_\textit{l}$ & Segment $l$ \\\hline
		$\boldsymbol{c}_j$ & Coordinate vector at the $j$th position \\\hline
		$\mathcal{R}$ & Road positioning space\\\hline
		$\mathcal{S}_i$ & Segment positioning space within the road $\rm{r}_\textit{i}$\\\hline
		$\mathcal{C}_l$ & Coordinate positioning space within the segment $\rm{s}_\textit{l}$\\\hline   		 
		$\boldsymbol{\phi}_i$ & HetNet signal sequence of the road $\rm{r}_\textit{i}$ \\\hline
		$\boldsymbol{o}_j$ & HetNet signal vector at the $j$th position \\\hline
		$P_{j,k}$ & RSS of the $k$th BS at the $j$th position \\\hline
		$g_{j,k}$ & Signal gradient of the $k$th BS at the $j$th position \\\hline
		$x_j$ & Longitude at the $j$th position \\\hline
		$y_j$ & Latitude at the $j$th position \\\hline 
		$\beta_k$ & Path loss parameter of the $k$th BS\\\hline
		$\boldsymbol{c}_{BS_k}$ & Coordinate vector of the $k$th BS position \\\hline
		$X_k$ & Noise of the $k$th BS\\\hline
		$\boldsymbol{f}_{\rm{r}_\textit{i}}$ & SF vector of the road $\rm{r}_\textit{i}$\\\hline
		$\boldsymbol{f}_{\rm{s}_\textit{l}}$ & SF vector of the segment $\rm{s}_\textit{l}$\\\hline
		$\boldsymbol{p}$ & Multi-scale position vector\\\hline 		
	\end{tabular}
	\vspace {-1 mm}
\end{table}
\subsection{Problem Formulation}
%定位问题用的是距离，实时搜索时用的是概率，怎么能结合起来，紧密相连。
To achieve low latency and high precision performance, the multi-scale vehicle localization problem can be formulated as follows:
\begin{alignat}{2}	
	\label{locproblem}
	\vspace{-1mm}
	\hat{\boldsymbol{p}} = & \begin{array}{*{20}{c}}
		\mathop{\arg\max}\limits_{{\rm{r}_\textit{i}},{\rm{s}_\textit{l}},{\boldsymbol{c}_j}} \quad &\omega_1\rm{Pr}({\rm{r}_\textit{i}})\rm{Pr}(\hat{\rm{r}_\textit{i}}\mid{\rm{r}_\textit{i}})\rm{Pr}({\rm{s}_\textit{l}} \mid \hat{\rm{r}_\textit{i}})\rm{Pr}(\hat{\rm{s}_\textit{l}}\mid{\rm{s}_\textit{l}}) \\
		&\cdot \rm{Pr}({\rm{c}_\textit{j}} \mid \hat{\rm{s}_\textit{l}})+(1-\omega_1)(\frac{O}{V})^{-1}
	\end{array} & \\ 
	\mbox{s.t.}\quad
	& {\rm{r}_\textit{i}} \in \mathcal{R}, {\rm{s}_\textit{l}}  \in {\mathcal{S}_{i}}, \boldsymbol{c}_{j} \in {\mathcal{C}_{l}}, & \tag{\ref{locproblem}a}\\
	& \omega_1 \in \left[0,1\right],& \tag{\ref{locproblem}b}
	\vspace{-2mm}
\end{alignat}
where $\rm{Pr}({\rm{r}_\textit{i}})\rm{Pr}(\hat{\rm{r}_\textit{i}}\mid{\rm{r}_\textit{i}})\rm{Pr}({\rm{s}_\textit{l}} \mid \hat{\rm{r}_\textit{i}})\rm{Pr}(\hat{\rm{s}_\textit{l}}\mid{\rm{s}_\textit{l}})\rm{Pr}({\rm{c}_\textit{j}} \mid \hat{\rm{s}_\textit{l}})$ denotes the joint Bayes probability of road matching, segment matching, and coordinate estimation. The ratio $\frac{O}{V}$ serves as an indicator of computation latency, where $O$ represents the number of CPU cycles executing the positioning task, which relies on the complexity of the positioning algorithm. Meanwhile, $V$ denotes the local computing rate of the VUE, quantified as the number of CPU cycles executed per second. In (\ref{locproblem}), $\omega_1$ is defined as the weighting factor of localization precision. The weighting factor can be improved for higher positioning accuracy. By optimizing the positioning problem in (\ref{locproblem}), we aim to minimize the computation latency and maximize the joint Bayes probability of the multi-scale positioning process. Hence, to solve this multi-scale vehicular positioning problem, three key sub-problems need to be solved: how to effectively divide segments with unequal lengths of different roads, how to reduce feature dimension to sparsely and accurately represent multi-scale positions, and how to quickly locate multi-scale positions with high precision.    
A\section{HetNet Signal Feature Extraction}
\label{Offline}
%概括说明本章的主要内容，研究内容与问题建模的联系
In this section, a SP segmentation method and a SF extraction method are proposed to respectively solve the sub-problems of partitioning segments and extracting sparse signal features. Specifically, to construct the segment positioning space, a SP segmentation method is first proposed by partitioning HetNet signal sequences based on signal gradients. Then, a SF extraction method is proposed to represent roads and segments in a sparse way. Meanwhile, the CRLB is derived to verify the positioning accuracy improvement brought from the segment partition and SF extraction. Additionally, to obtain vehicular locations within segments, HetNet signals are mapped to coordinates by curve fitting.   
\subsection{HetNet Signal Sequence Segmentation}\label{segprocess}
%特征分段，先说明多尺度存在性，给出地图和实测信号的梯度变化图，说明用户的移动轨迹存在道路、子段、坐标这种尺度变化关系
%然后说道路划分和子段划分的准则，依据什么划分，为什么，子段给出CRLB的推导说明用梯度平方和特征可以降噪，且说明分段点选择方法
\subsubsection{SP Extraction and Sequence Segmentation}\label{segmentation method}
Segment positioning refers to the process of determining the localized segment of the road on which a vehicle is moving. To construct segment positioning space, HetNet signal sequence partition is crucial. Nevertheless, balancing the partitioning lengths and the segment positioning accuracy is challenging. To partition segments with adaptive lengths, gradient features can be applied for partitioning adjacent road sections since signal gradients will change regularly in LoS and NLoS links in different road sections caused by fixed regional obstacles \cite{5767589}, e.g., trees, buildings, bridges, etc. To accurately partition each road based on signal gradient features, the SP is defined, and then a segmentation method is designed by extracting SPs of HetNet signal sequences.
\par
For ease of analysis, we give the definitions of SPs as follows: 

\begin{definition}[\textbf{SP}]
	SPs are defined as HetNet signal vectors employed to partition the roads into segments as (\ref{singulardefination}), where the signal gradient symbol or value of at least one BS changes at each SP as qualified in (\ref{gradientdifferencedefination}).   
\end{definition}
\begin{equation}
\begin{aligned}\label{singulardefination}
	%	\vspace{-1mm}
	%		\setlength\abovedisplayskip{3pt}		
	%		\vspace{-5mm}
	\boldsymbol{\phi}_i = &\underbrace{\left[\boldsymbol{o}_1, \boldsymbol{o}_2, \hdots, \boldsymbol{o}_L\right]^\top}_{\rm{Road} \thinspace \rm{r}_\textit{i}}
	= [ \underbrace{[\boldsymbol{o}_1,\hdots, {\overbrace{\boldsymbol{o}_{b_1}}^{\rm{SP_1}}} ]}_{\rm{Segment} \thinspace \rm{s}_1},\\
	&\underbrace{[\boldsymbol{o}_{b_1+1},\hdots, \overbrace{\boldsymbol{o}_{b_2}}^{\rm{SP_2}}]}_{\rm{Segment} \thinspace \rm{s}_2}, \hdots, \underbrace{[\overbrace{\boldsymbol{o}_{b_{{n_i}}}}^{\rm{SP}_{\textit{n}_i}},\hdots, \boldsymbol{o}_{L_i} ]}_{\rm{Segment} \thinspace \rm{s}_{\textit{n}_i+1}}]^\top,
\end{aligned}
\end{equation}
where $\boldsymbol{\phi}_i$ is the HetNet signal sequence collected from the road $\rm{r}_\textit{i}$. A segment is represented by a HetNet signal sequence, where the endpoints of this sequence are SPs. In (\ref{singulardefination}), $b_j$ ($b_j \in \left[1, {L_i}\right]$) is the position index of the SP, where $L_i$ is the position number within the road $\rm{r}_\textit{i}$. At the $b_j$th signal sampling position, the SP is denoted by the vector of HetNet RSSs, i.e., $\boldsymbol{o}_{b_j} = \left[P_{b_j,1}, P_{b_j,2},\hdots, P_{b_j, K}\right]^\top$. In  (\ref{singulardefination}), $n_i$ denotes the number of SPs that varies with roads. Since SPs essentially represent key change points of the signs or the values in gradient features, the length of a SP set $n_i$ is determined by the number of extreme points of a HetNet signal sequence and the change threshold of gradient values. To ensure that there exist $n_i$ SPs to partition a road into multiple segments as (\ref{singulardefination}), $n_i$ fulfills the condition $n_i \in \left[0, L_i-2\right]$, where $n_i = 0$ denotes that there is no gradient change of any BS signals within the road, and $n_i = L_i-2$ indicates that the signal gradient difference on both sides of each signal sampling position on the road is greater than the threshold. To accurately describe the sign or the value change of the gradient feature, for $\forall b_j \in \left[1, {L_i}\right], \forall k \in \left[1, K\right]$, the signal gradient at each SP is constrained as follows:
\begin{equation}\label{gradientdifferencedefination}
	%	\vspace{-1mm}
	\setlength\abovedisplayskip{3pt}
	\begin{cases}
		\mathrm{sgn} \left(g_{b_j+1,k} \cdot g_{b_j,k}\right) \textless 0, \\
		or,\\
		\vert g_{b_j+1,k} - g_{b_j,k} \vert \geq \tau,
	\end{cases}		
	\vspace{-1mm}
\end{equation}
where $g_{b_j+1,k}$ and $g_{b_j,k}$ respectively denote the signal gradients of the $k$th BS on two adjacent positions, and $\tau$ represents the gradient change threshold. In (\ref{gradientdifferencedefination}), $\mathrm{sgn} \left(g_{b_j+1,k} \cdot g_{b_j,k}\right) \textless 0$ and $\vert g_{b_j+1,k} - g_{b_j,k} \vert \geq \tau$ respectively determine the sign and value changes of the gradient at a SP. By the gradient constraint in  (\ref{gradientdifferencedefination}), the gradient sign or the gradient value of at least one BS signal changes. Based on the SP definition in (\ref{singulardefination}) and (\ref{gradientdifferencedefination}), the position index set of all SPs within a road-scale HetNet signal sequence $\boldsymbol{\phi}_i$ is denoted as $\mathcal{B}_i=\left\{ b_1, b_2,..., b_{{n_i}} \right\} $. This position index set $\mathcal{B}_i$ can be used to extract all SPs within a road.
\par
Due to the varying lengths and shapes of roads, the number of SPs also differs, resulting in segments with different lengths. To accurately partition different roads, we propose a SP segmentation method. This method first extracts the position index set $\mathcal{B}_i$, and then partition this road based on the SPs of $\mathcal{B}_i$. Specifically, to improve the anti-noise performance and avoid the cancellation of signals at adjacent positions during the calculation of the average gradient, the signed mean square gradient feature is employed to partition roads. Based on the signed mean square gradient features, the sequence segmentation method can be designed by extracting the position index set of SPs as follows:	
\vspace{-1mm}
\begin{equation}\label{segproblem}	
	\mathcal{B^\ast} = \mathop{\arg\min}\limits_{\mathcal{B}} 
	{\sum\limits_{k = 1}^{K} \sum\limits_{j = {b_l}}^{b_{l+1}}} \frac{\left({\mathrm{sgn}}(P_{j+1,k}-P_{j,k}) {g_{_{j,k}}^2}- \overline{g_{{{b_l}\cdot{b_{l+1}}},k}^{_2}} \right)^2}{b_{l+1}-b_l+1}, 
    \vspace{-1mm}
\end{equation}
where $K$ is the number of all HetNet BSs, and $g_{_{j,k}}$ denotes the signal gradient feature as (\ref{g}). ${\mathrm{sgn}}(P_{j+1,k}-P_{j,k})$ represents the sign function which is utilized to determine the sign of the gradient at adjacent positions as follows:
\begin{equation}
	\mathrm{sgn}(P_{j+1,k}-P_{j,k}) =
	\left \{
	\begin{aligned}
		\; 1 & \quad & P_{j+1,k}-P_{j,k} > 0 \\
		\; 0 & \quad & P_{j+1,k}-P_{j,k} = 0 \\
		\; -1  & \quad & P_{j+1,k}-P_{j,k} < 0
	\end{aligned}.
	\right.
\end{equation}
$\overline{g_{{{b_l}\cdot{b_{l+1}}},k}^{_2}}$ denotes the average value of the signed mean square gradient of the $k$th BS between the $b_l$th position and ($b_l+1$)th position, which can be expressed as follows:
\begin{equation}\label{meangradient}
	\vspace{-3mm}
	\overline{g_{{{b_l}\cdot{b_{l+1}}},k}^{_2}}  = \frac{1}{b_{l+1}-b_l+1}\sum\limits_{j = b_l}^{b_{l+1}} {\mathrm{sgn}}(P_{j+1,k}-P_{j,k}){g_{j,k}^2 }.		
	\vspace{1mm}
\end{equation}
Extracting $\mathcal{B^\ast}$ is highly non-trivial since SPs determine the length of the partitioned segments and the accuracy of the segment features. To extract the optimal position index set $\mathcal{B^\ast}$ within the HetNet signal sequence $\boldsymbol{\phi}_i$, the Bottom-up algorithm \cite{TRUONG2020} is utilized.  
\par
\begin{figure}
	\centering
	\scriptsize
	\begin{tabular}{cc}		
		
		\subfigure [HetNet BS RSS values.] {
			\vspace{3 mm}
			\label{figure:subfig_a}			
			\centering
			%		\vspace{1 mm}
			\includegraphics [width=4.5cm]{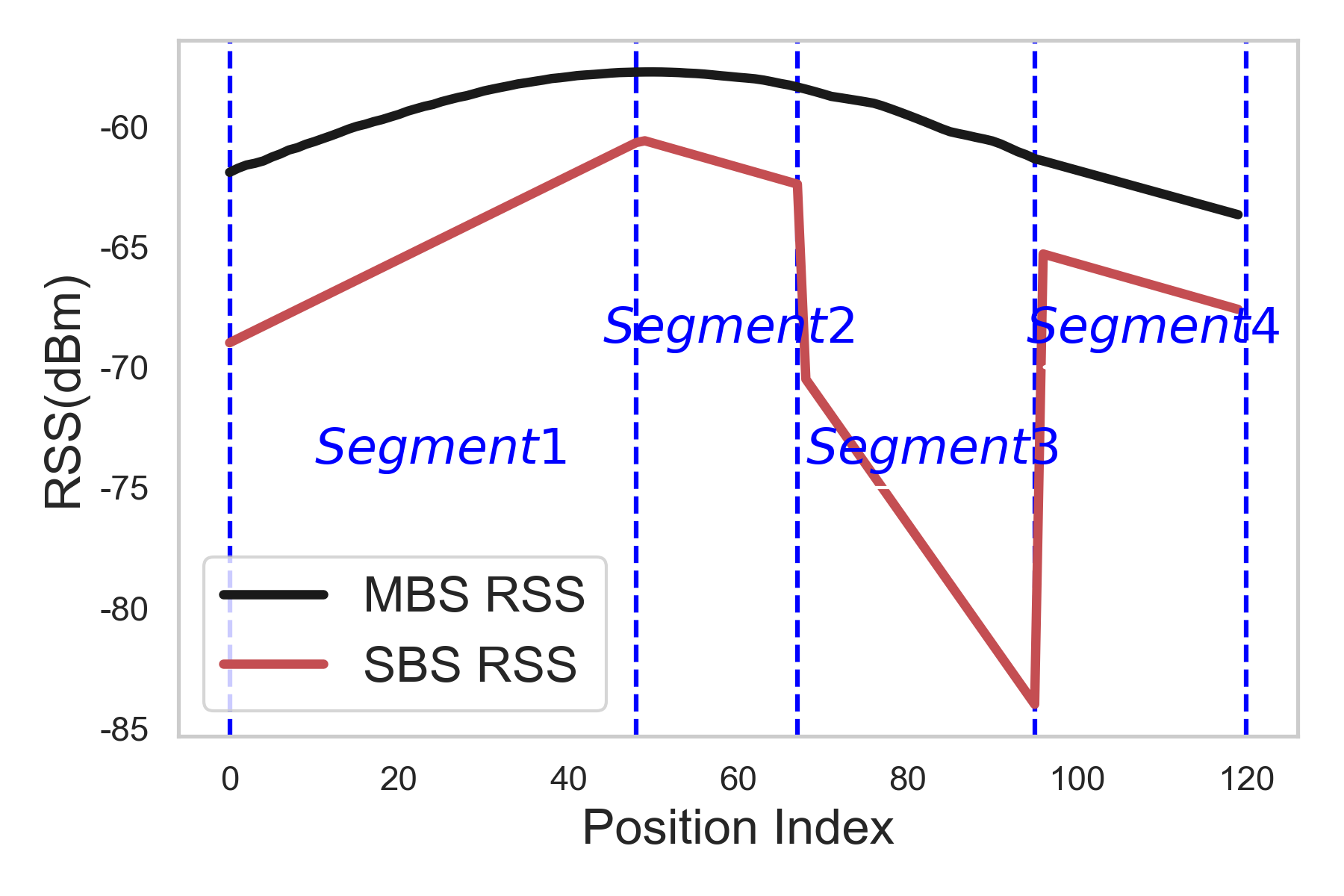}					
		} \\
		\hspace{-7mm}  		
		\subfigure [Signed mean square gradient feature.] {
			\vspace{3 mm}
			\label{figure:subfig_b}
			\centering
			\includegraphics[width=4.5cm]{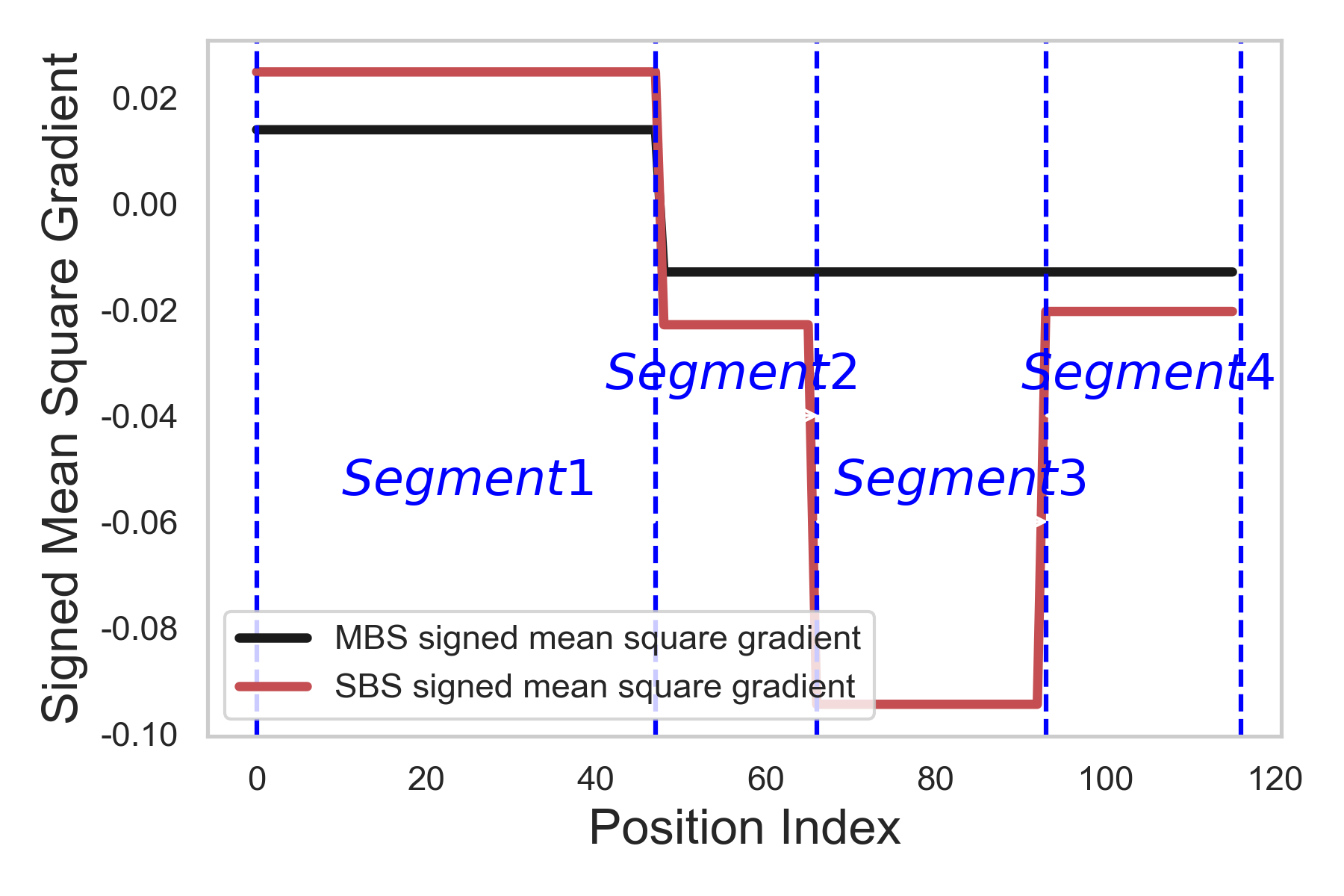}}\hspace{-5mm}  		
		\subfigure [Mean RSS feature.] {
			\vspace{3 mm}
			\label{figure:subfig_c}
			\centering
			\includegraphics[width=4.5cm]{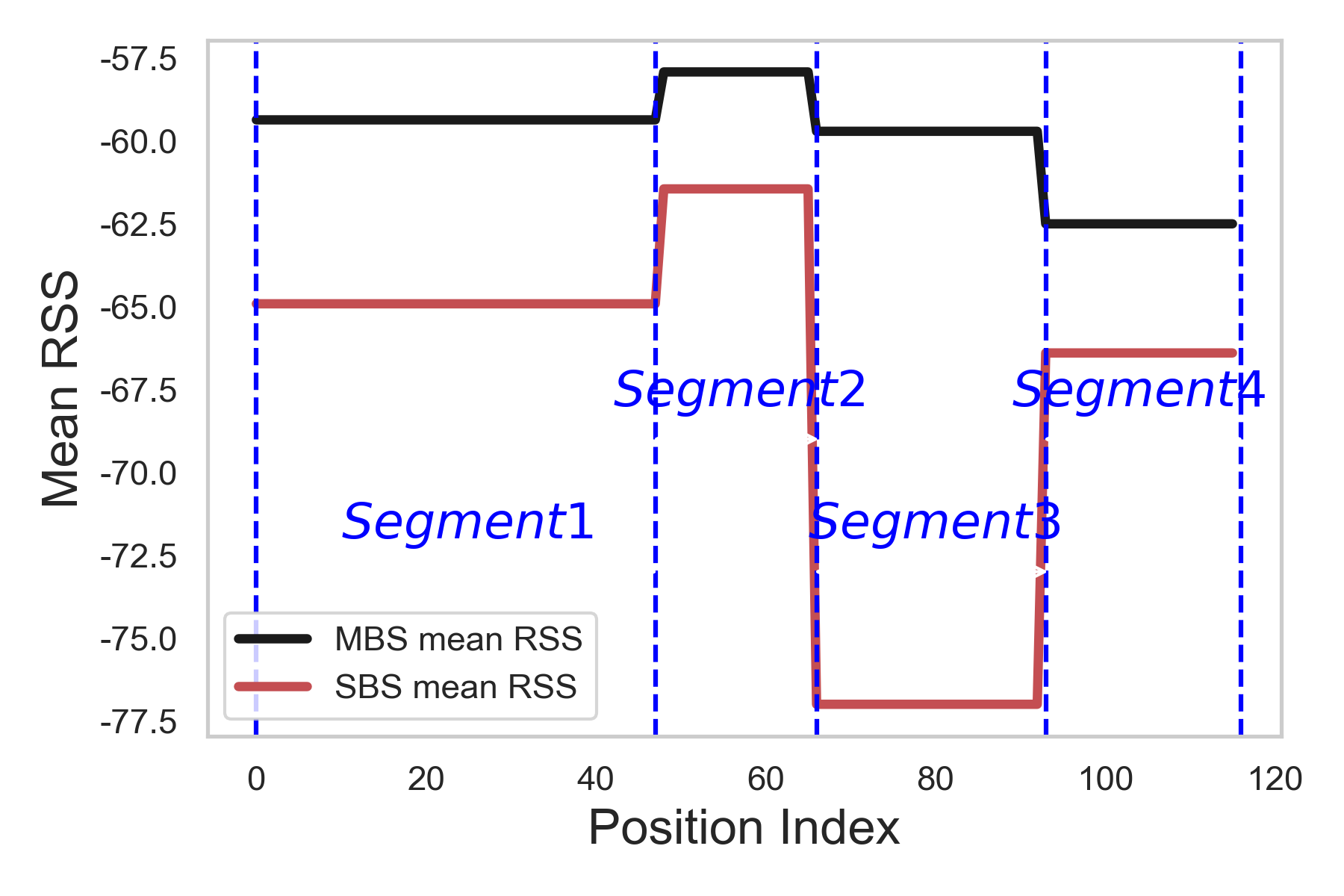}}\hspace{-5mm}		
		\centering
	\end{tabular}
	\vspace{-3 mm}
	%\captionsetup{labelfont={color=red}}
	\caption{An example of HetNet signals along a road.}
	\label{figure}
	\vspace{-5 mm}
\end{figure}

To demonstrate that our proposed method can effectively extract SPs that satisfy their definitions in (\ref{singulardefination}) and (\ref{gradientdifferencedefination}), the SP partitioned segment-scale HetNet signal sequences and corresponding gradient signal features are shown in Fig. \ref{figure}. If all $b_l$ and $b_{l+1}$ are extracted according to Fig. \ref{figure:subfig_b}, the optimal solution for segment 1's SPs position indexes is 1 and 48. However, if 1 and 60 are chosen as the SFs of segment 1, the gradient variance in (\ref{segproblem}) will increase, making it challenging to achieve the goal of minimizing the sum of gradient variance. Therefore, the extracted SPs of (\ref{segproblem}) can meet the gradient constraint of SP definition in (\ref{gradientdifferencedefination}), and road-scale HetNet signal sequence can be automatically partitioned into segments with different gradients.
\par
Based on two adjacent position indexes of the extracted SPs, a road-scale HetNet signal sequence $\boldsymbol{\phi}_i$ can be partitioned into multiple segment-scale HetNet RSS sequences, i.e., $\hat{\boldsymbol{\phi}_i}=\left\{ \boldsymbol{\psi}_1,\hdots,\boldsymbol{\psi}_{n_i+1} \right\}$, where  $\boldsymbol{\psi}_l=\left\{\boldsymbol{o}_{b_l},\hdots,\boldsymbol{o}_{b_{l+1}}\right\}$, and $n_i+1$ denotes the number of partitioned segment-scale HetNet RSS sequences. Corresponding to segment-scale HetNet signal sequences in $\hat{\boldsymbol{\phi}_i}$, segments within road $\rm{r}_\textit{i}$ can be partitioned, and these segments are defined as a segment space $\mathcal{S}_i=\left\{ {\rm{s}_1},{\rm{s}_2},\hdots,{\rm{s}}_{n_i+1} \right\}$, where $\rm{s}_\textit{l}$ denotes the $l$th segment on the road $\rm{r}_\textit{i}$.
\subsubsection{Segmentation Bound Derivation and Proof}
To analyze positioning errors of different segmentation bounds, the CRLB of segment positioning is derived. Based on the signed mean square gradient $\overline{{g_{j}^{l}}^2}$ of the $l$th segment, the CRLB takes the form
\begin{equation}\label{CRLBseg}
	\vspace{-1mm}
	tr({\rm{CRLB}}) \geqslant \frac{{\ln^2{10}}}{\frac{{4\cdot10^4}}{\left({N_l}-1\right)^2}{\sum_{k=1}^{K}\frac{{\beta_k^l}^4}{{\rho_k^l}^2}(\frac{1}{d_{{N_l},k}^l}-\frac{1}{d_{1,k}^l}\frac{1}{N_l})^2}},		
	\vspace{-1mm}
\end{equation} 
where $\beta_k^l$ and ${\rho_k^l}^2$ respectively represent the path-loss exponent and the RSS noise covariance of the $k$th BS within the $l$th segment. The endpoints of a segment are defined as the beginning and the end HetNet RSS vectors of this segment-scale HetNet signal sequence. In (\ref{CRLBseg}), $d_{N_l,k}^l$ and $d_{1,k}^l$ represent signal propagation distances of the $l$th segment endpoints, which are decided by the extracted SPs of the proposed sequence segmentation method. $N_l$ denotes the number of signal sampling positions within the $l$th segment. The detailed CRLB calculation of (9) is presented in Appendix A.
\par
If the signal propagation environments within a segment remain stable, the derivation of the CRLB in (\ref{CRLBseg}) demonstrates that the minimum achievable positioning error is intrinsically linked to the signal propagation distances at the points where the segment is partitioned. Based on the path-loss model in (\ref{plm}), the RSS values can be utilized to calculate corresponding signal propagation distances. To achieve the lowest CRLB, ${P_{N_l,k}^{l}}'$ and ${P_{1,k}^{l}}'$ respectively represent the RSS values of the $k$th BS at the endpoints of the $l$th segment RSS sequence. Meanwhile, $P_{max,k}^l$ and $P_{min,k}^l$ represent the extreme points, which are the maximum and minimum RSSs of the $k$th BS signals within the $l$th segment RSS sequence. The difference between endpoints and extreme points of segment RSS sequence is that endpoints represent the beginning and the end points which can be any RSS values, but the extreme points only denote the maximum and minimum RSS values within the HetNet signal sequence. The CRLB of the segment partitioned by the RSS extreme points is expressed as ${\rm{CRLB}}_{P'}$, and ${\rm{CRLB}}_{P}$ represents the CRLB of the segment that is not divided by the RSS extreme points. Therefore, Proposition \ref{CRLBsegpo} is proposed to prove that the segments approaching the lowest CRLB can be obtained by the SPs that are the extreme points of a road-scale HetNet signal sequence.   
\newtheorem{proposition}[theorem]{\textbf{Proposition}}
\begin{proposition}\label{CRLBsegpo}
	With the same path-loss exponent $\beta_k^l$, if the beginning point and the end point of one segment-scale signal sequence from the $k$th BS respectively satisfies ${P_{N_l,k}^l}' =  P_{max,k}^l$ and ${P_{1,k}^l}'= P_{min,k}^l$, ${\rm{CRLB}}_{P'} < {\rm{CRLB}}_{P}$. 
\end{proposition}
\par
\begin{proof}
	Please refer to Appendix B.
\end{proof}
\par
As derived in Proposition \ref{CRLBsegpo}, when the endpoints of a segment HetNet signal sequence are the extreme points of the road-scale HetNet RSS sequence, the segment can be accurately located approaching the lowest positioning error. 
\par
Since the path-loss exponent $\beta_k^l$ can influence RSS measurements, it is important to investigate the gradient turning points, which are the other type of SPs, to ensure an accurate segment partition. To this end, we propose the Remark \ref{segbeta} to analyze the impact of signal propagation environments on segment partitioning.  
\begin{remark}\label{segbeta}
	 To approach the lowest CRLB, a segment signal sequence can be further partitioned based on the gradient turning points of the $k$th BS signals if the $\beta_k^l$ changed within a segment. 
\end{remark}
\par
Based on the Remark \ref{segbeta}, the signal path-loss exponent $\beta_k^l$ is stable within the segmented signal sequences. In this case, road-scale signal sequences can be partitioned into segment signal sequences approaching the lowest CRLB as (\ref{CRLBseg}).   

\subsection{Signal Feature Extraction and Representation}
%特征提取，先说道路需要表征，为了解决道路表征先提出多特征的特征集，说明这样表征的CRLB低，然后进一步为降低延迟，remark定理给出，引出特征选择的方法
Due to the high-dimensional signals in the segmented HetNet signal sequences, a feature set composed of signal statistical features can be extracted to represent roads and segments. However, not all features have obvious distinctions between different road-scale signal sequences. If the signal features with significant differences could be extracted for feature matching, not only would the matching accuracy be improved, but also would the computation complexity be reduced. To this end, the road and segment representation consists of three processes, namely feature set extraction, SF extraction, and sparse feature representation.  
\subsubsection{\textbf{Feature Set Extraction}}
Representing long roads by a single gradient feature is challenging since complex signal fluctuation trends exist along roads caused by different signal attenuation strengths of HetNet signals. Hence, it is critical to represent roads with multiple signal features to improve the representation accuracy. Then, using what kind of features is beneficial for our proposed mechanism? Since RSS values of HetNet signals can be obtained without additional information on BS locations and transmission powers, statistical signal features are extracted from HetNet RSSs, which can save the storage space and alleviate the computational burden on resource-limited mobile devices. Moreover, as the received signal power is essentially obtained by the average amplitude of different frequency sub-carriers, the RSS level can reflect the amplitude feature of the signal's frequency domain and indicate path loss related to the environment and distance. Therefore, by extracting RSS features from multiple BSs, the adverse effects on positioning accuracy can be mitigated though there are frequency domain feature fluctuations of a single BS. To accurately reflect the signal features of different roads, various statistical features of HetNet RSSs can be extracted to form a feature set. 
\par
To characterize high-dimensional signal data of roads by only a few signal features, a feature set is extracted including the statistical signal features and correlation signal features. To avoid the impact of different environmental conditions on signal attenuation, feature sets can be extracted separately for different weather conditions. In this paper, sunny climate is selected as a typical weather extracting signal features and designing localization algorithm. To this end, under a clear weather, a feature set extracted from a road-scale HetNet signal sequence can be defined as $\boldsymbol{e}_{\rm{r}_\textit{i}}$, which is expressed as follows:  
\begin{equation}\label{featureset}
	%	\vspace{-1mm}
	\setlength\abovedisplayskip{3pt}
	\setlength\belowdisplayskip{3pt}
	\begin{split}
		\boldsymbol{e}_{\rm{r}_\textit{i}}= \left[\overline{g_{1}^{_2}}, \hdots, \overline{g_{K}^{_2}}, {\mu}_1,\hdots, {\mu}_K,{\sigma}_1,\hdots, \right. \\	
		\left.  {\sigma}_K, \overline{\Delta}_1,\hdots, \overline{\Delta}_K, \chi_1,\hdots,\chi_K \right]^\top.
	\end{split}
	% 	\vspace{-1mm}
\end{equation}	
The feature number of the $\boldsymbol{e}_{\rm{r}_\textit{i}}$ is $qK$, where $q$ represents the number of feature kinds. The feature descriptions of the $K$th BS signals in $\boldsymbol{e}_{\rm{r}_\textit{i}}$ are shown as Table \ref{tab:tab2}. 
\newcommand{\tabincell}[2]{\begin{tabular}{@{}#1@{}}#2\end{tabular}}
\begin{table}[h]	
	\centering	
	\caption{Feature Description }\label{tab:tab2}
	\vspace{-2mm}
	\begin{tabular}{|c || c|}
		\hline
		Notation & Physical meaning\\\hline 
		$\overline{g_{K}^{_2}}$ & {The signed mean square gradient feature } \\\hline 
		${\mu}_K$ & \tabincell{c} {The mean RSS feature} \\\hline
		${\sigma}_K$ & \tabincell{c} {The RSS variance feature} \\\hline   		 
		$\overline{\Delta}_K$ & \tabincell{c} {The average difference between two BS RSSs} \\\hline 
		$\chi_K$ & \tabincell{c} {The RSS range feature}\\\hline 		
	\end{tabular}
	\vspace {-1 mm}
\end{table}
\par
To theoretically analyze that localization errors can be reduced by matching multiple features, the CRLBs under different feature numbers are derived. The CRLB of one signal feature, i.e., the signed mean square gradient feature $\overline{g^{_2}}$, is denoted as ${\rm{CRLB}}_{\overline{g^{_2}}}$, while ${\rm{CRLB}}_{\mu,\overline{g^{_2}}}$ represents the CRLB of two signal features, i.e., the signed mean square gradient feature $\overline{g^{_2}}$ and the mean RSS feature $\mu$.     
\begin{proposition}\label{CRLBfeatureset}
	In the segment positioning process, the error of matching multiple features is smaller than the error of matching a single gradient feature, i.e., ${\rm{CRLB}}_{\mu,\overline{g^{_2}}}<{\rm{CRLB}}_{\overline{g^{_2}}}$. 
\end{proposition}

\begin{proof}
	Please refer to Appendix C.
\end{proof}
\par
As derived in Proposition \ref{CRLBfeatureset}, the CRLB only using one signal feature, i.e., the signed mean square gradient, is $tr({\rm{CRLB}}_{\overline{g^{_2}}}) \geqslant \frac{{\ln^2{10}}}{\frac{{4\cdot10^4}}{\left(N_l-1\right)^2}{\sum_{j=1}^{K}\frac{\beta_j^4}{\rho_j^2}(\frac{1}{d_{N_l,j}}-\frac{1}{d_{1,j}}\frac{1}{N_l})^2}}$. Supposed that the mean RSS feature exists differences among segments, the CRLB with both the mean square gradient and mean RSS is $tr({\rm{CRLB}}_{\overline{g^{_2}},\mu}) \geqslant \frac{4\ln^2{10}}{{\frac{{4^2\cdot10^4}}{\left(N_l-1\right)^2}{\sum_{j=1}^{K}\frac{\beta_j^4}{\rho_j^2}(\frac{1}{d_{N_l,j}}-\frac{1}{d_{1,j}}\frac{1}{N_l})^2}}+{2\frac{20^2}{N_l^4\cdot}}{\sum_{j=1}^{K}\frac{\beta_j^2}{\eta_j^2}(\sum_{i=1}^{N_l}\frac{i}{d_{ij}})^2}}$. Hence, the CRLB difference is $tr({\rm{CRLB}}_{\overline{g^{_2}}})-tr({\rm{CRLB}}_{\overline{g^{_2}},\mu}) \geqslant \frac{4b}{a(a+b)}$, where $a=\frac{{4^2\cdot10^4}}{\left(N_l-1\right)^2{\ln^2{10}}}{\sum_{j=1}^{K}\frac{\beta_j^4}{\rho_j^2}(\frac{1}{d_{N_l,j}}-\frac{1}{d_{1,j}}\frac{1}{N_l})^2}$ and $b={2\frac{20^2}{N_l^4\cdot\ln^2{10}}}{\sum_{j=1}^{K}\frac{\beta_j^2}{\eta_j^2}(\sum_{i=1}^{N_l}\frac{i}{d_{ij}})^2}$. Since $\frac{4b}{a(a+b)}>0$, positioning accuracy can be improved by the utilization of both the mean square gradient and mean RSS features. Therefore, by incorporating the additional information from the mean RSS, the CRLB can be further reduced, resulting in a localization error reduction of up to $\frac{4b}{a(a+b)}$. Meanwhile, the complexity of positioning computations can be reduced by utilizing the feature set vector to represent a road since the feature number $qK$ is much smaller than the number of signal sampled positions $N_l$, i.e., $qK \ll N_l$. The theoretical analysis mentioned above can be intuitively illustrated in Fig. \ref{figure}. As depicted in Fig. \ref{figure:subfig_b}, the road can be partitioned into four segments, where SBS's signed mean square gradient is different between two adjacent segments. However, similar gradients may exist between two non-adjacent segments, e.g., the segment 2 and the segment 4. By matching the different mean RSS features as shown in Fig. \ref{figure:subfig_c}, the non-adjacent segments with similar gradients can be localized. Therefore, more statistical features are essential to identify these segments with similar gradient features to ensure segment positioning accuracy. 
 
\subsubsection{\textbf{SF Extraction}}
Since not all features have significant distinctions between positions within the same spatial scale, signal features with obvious distinctions should be extracted for position representation to reduce feature-matching complexity and localization latency. By comparing the signed mean square gradient and mean RSS of the MBS in segments 2, 3, and 4 in Fig. \ref{figure:subfig_b} and Fig. \ref{figure:subfig_c}, it is obvious that the mean RSS feature can be extracted as a salient feature distinguishing  these segments. The reason is that the signed mean square gradient varies little between different segments, while the mean RSS feature shows a significant difference.
\par
To prove that the feature-matching complexity can be further reduced without compromising localization accuracy by matching SFs, Proposition \ref{featurereduce} is proposed. ${\rm{CRLB}}_{K}$ denotes that the CRLB is derived based on $K$ signal features, and ${\rm{CRLB}}_{K-1}$ represents that $K-1$ signal features are utilized for deriving the CRLB.   
\begin{proposition}\label{featurereduce}
	In the positioning space of roads and segments, if the normalized signal features of the $K$th BS changed within $\left[-\gamma,\gamma\right]$ between all adjacent positions, the CRLB does not change without matching the signal feature of the $K$th BS. Specifically, ${\rm{CRLB}}_{K}={\rm{CRLB}}_{K-1}$.
\end{proposition}
\begin{proof}
	Please refer to Appendix D.
\end{proof}
\par
As derived in Proposition \ref{featurereduce}, by matching the SF with redundant information removed, the computational complexity can be reduced and the positioning accuracy can be ensured at the same time. Therefore, it is essential to design a method to extract SFs for quickly locating vehicles.  
\par
To extract distinctive features for position representation, we propose a SF extraction method by maximizing information gain. Specifically, a feature set vector $\boldsymbol{e}_{\rm{r}_\textit{i}}$ in (\ref{featureset}) is first extracted from a road-scale signal sequence $\boldsymbol{\phi}_i$. The potential SFs of road $\rm{r}_\textit{i}$ is expressed as a vector $\boldsymbol{f}_{\rm{r}_\textit{i}}$, which is a subset of the feature set $\boldsymbol{e}_{\rm{r}_\textit{i}}$. Then, to extract the SF vector $\boldsymbol{f}_{\rm{r}_\textit{i}}^\ast$ from $\boldsymbol{e}_{\rm{r}_\textit{i}}$, a SF extraction method is designed as follows:  
\begin{equation}\label{IC}
	\setlength\abovedisplayskip{2pt}
	\setlength\belowdisplayskip{2pt}
	\boldsymbol{f}_{\rm{r}_\textit{i}}^\ast =\mathop{\arg\max}_{\boldsymbol{f}_{\rm{r}_\textit{i}} \subseteq \boldsymbol{e}_{\rm{r}_\textit{i}}}
	H(\rm{r}_\textit{i})-H\left({\rm{r}_\textit{i}} \mid {\boldsymbol{f}_{\rm{r}_\textit{i}}}\right),
	%	\vspace{-1mm}
\end{equation}
where $H(\rm{r}_\textit{i})$ denotes the road information entropy of RSS values in $\boldsymbol{\phi}_i$, and $H\left({\rm{r}_\textit{i}} \mid {\boldsymbol{f}_{\rm{r}_\textit{i}}}\right)$ represents the road conditional entropy of the SF vector $\boldsymbol{f}_{\rm{r}_\textit{i}}$. The conditional entropy in (\ref{IC}) is expressed as follows:
\begin{equation}\label{Hcon}
	\vspace{-1mm}
	H\left({\rm{r}_\textit{i}} \mid {\boldsymbol{f}_{\rm{r}_\textit{i}}}\right)= 
	-\sum\limits_{j = 1}^{N_{\boldsymbol{f}_{\rm{r}_\textit{i}}}}\sum\limits_{i = 1}^{m}
	p\left({\rm{r}_\textit{i}},\boldsymbol{f}_j\left(\boldsymbol{\phi}_i\right)\right)\log{p\left({\rm{r}_\textit{i}}\mid {\boldsymbol{f}_j\left(\boldsymbol{\phi}_i\right)}\right)},
	\vspace{1mm}
\end{equation}
where $p\left({\rm{r}_\textit{i}},\boldsymbol{f}_j\left(\boldsymbol{\phi}_i\right)\right)$ represents the joint probability of road ${\rm{r}_\textit{i}}$ and the $j$th feature in the feature set $\boldsymbol{e}_{\rm{r}_\textit{i}}$, and $N_{\boldsymbol{f}_{\rm{r}_\textit{i}}}$ denotes the number of feature kinds within the SF vector $\boldsymbol{f}_{\rm{r}_\textit{i}}$. The number of roads in the road positioning space is represented by $m$. $p\left({\rm{r}_\textit{i}}\mid {\boldsymbol{f}_j\left(\boldsymbol{\phi}_i\right)}\right)$ in (\ref{Hcon}) is the conditional probability of road ${\rm{r}_\textit{i}}$ based on the given feature subset ${\boldsymbol{f}_j\left(\boldsymbol{\phi}_i\right)}$, and is expressed as
\begin{equation}\label{subfeature}
	\vspace{-1mm}
	p\left({\rm{r}_\textit{i}}\mid {\boldsymbol{f}_j\left(\boldsymbol{\phi}_i\right)}\right)=
	\frac{p\left({\rm{r}_\textit{i}}\right){p\left({\boldsymbol{f}_j\left(\boldsymbol{\phi}_i\right)}\mid {\rm{r}_\textit{i}}\right)}}{p\left(\boldsymbol{f}_j\left(\boldsymbol{\phi}_i\right)\right)}.
	\vspace{-1mm}
\end{equation} 
The optimal SF vector with the maximum information gain can be extracted by exhaustive searching. These extracted salient features can accurately reflect different signal propagation environments of the LoS and the NLoS links, achieving high-precision location mapping by low-dimensional features. Moreover, positioning errors caused by signal fluctuations at a single location can be mitigated by SFs since these features are extracted from multiple signal sampling positions. As the vehicle moves, the propagation environments of LoS and NLoS signals continuously change, resulting in the detection of new signals while older signals may be lost. By the proposed SF extraction method, the differences between detected signals are employed as beneficial factors that can be effectively utilized to enhance positioning accuracy.
\par
The extracted SFs are based on path loss and slow fading models, which can represent signal propagation distance and environment on a large scale, thereby enabling precise mapping of vehicle multi-scale locations. On the other hand, due to weather variations and the movement of obstacles, the signals may rapidly fluctuate on a small scale caused by fast fading effects. The rapid changes of these signals over short distances and short periods introduce additional noise, which can be mitigated in the following two ways. First, in the processes of extracting SFs, signal features from multiple BSs have been employed to avoid substantial fluctuations caused by fast-fading effects associated with a single BS. Second, the average of multiple signal measurements at a given location is utilized to obtain a reliable signal value, while conducting statistical feature analyses on the signals within a road region. Therefore, SFs extracted by our proposed method can filter noise from signal fluctuations, thereby ensuring positioning accuracy in the presence of fast fading.    
\subsubsection{\textbf{Sparse Feature Representation}}
Since the extracted SFs vary with different roads and segments, to uniformly represent these different SFs, a sparse representation model is designed as follows:
\begin{equation}\label{1}
	\setlength{\abovedisplayskip}{2pt}
	\boldsymbol{f}_{\rm{r}_\textit{i}}={\boldsymbol{W}_{\rm{r}_\textit{i}}}{\boldsymbol{e}_{\rm{r}_\textit{i}}},
	\vspace{-1mm}
\end{equation}
where $\boldsymbol{W}_{\rm{r}_\textit{i}} \in \mathbb{R}^{qK \times qK}$ represents a feature selection matrix that is denoted by a $qK \times qK$ diagonal matrix, and $qK$ represents the total number of features in the feature set vector $\boldsymbol{e}_{\rm{r}_\textit{i}}$. The subscript of $\boldsymbol{W}_{\rm{r}_\textit{i}}$ denotes that this feature extraction matrix corresponds to the road $\rm{r}_\textit{i}$. In (\ref{1}), $w_{kk} (k \in \left[1,qK\right])$ represents the selection coefficient of the $k$th feature, which is defined as a binary variable, i.e., $w_{kk} = 1$ if the $k$th feature of $\boldsymbol{e}_{\rm{r}_\textit{i}}$ is a SF. Otherwise, $w_{kk} = 0$. Since the SF vector can be represented by multiplying the feature selection matrix and the feature set vector, a vector composed of all road-scale SFs is defined as $\boldsymbol{f}_{\rm{r}}$, which is expressed as follows:
\begin{equation}\label{F_r}
	\setlength{\abovedisplayskip}{2pt}
	\boldsymbol{f}_{\rm{r}}=\left[\boldsymbol{f}_{\rm{r}_1},...,\boldsymbol{f}_{\rm{r}_m}\right],
	\vspace{-1mm}
\end{equation}
where $\boldsymbol{f}_{\rm{r}_\textit{i}}$ denotes the SF vector of the road $\rm{r}_\textit{i}$, and $\boldsymbol{f}_{\rm{r}_\textit{i}}$ is sparsely represented by ${\boldsymbol{W}_{\rm{r}_\textit{i}}}{\boldsymbol{e}_{\rm{r}_\textit{i}}}$. 
\par
Since similar gradients may exist in some non-adjacent segments, e.g., the segment 2 and the segment 4 in Fig. \ref{figure:subfig_b}, more statistical features with differences can be extracted to represent segments, e.g., mean RSS values showing in Fig. \ref{figure:subfig_c}. Therefore, to improve the representation precision, SFs of segments can be selected by the proposed SP extraction method. To this end, the SF vector $\boldsymbol{f}_{\rm{s}_\textit{j}}$ of the $j$th segment $\rm{s}_\textit{j}$ on the road $\rm{r}_\textit{i}$ can be extracted from corresponding feature set vector $\boldsymbol{e}_{\rm{s}_\textit{j}}$ through the proposed SF extraction method. Based on the proposed sparse representation model, the extracted $\boldsymbol{f}_{\rm{s}_\textit{j}}$ can be expressed as $\boldsymbol{W}_{\rm{s}_\textit{j}}\boldsymbol{e}_{\rm{s}_\textit{j}}$, where $\boldsymbol{W}_{\rm{s}_\textit{j}} \in \mathbb{R}^{L \times L}$ denotes the feature selection matrix of the $j$th segment $\rm{s}_\textit{j}$. Then, a vector composed of all segment-scale SFs is defined as $\boldsymbol{f}_{\rm{s}}$ that takes the form of:
\begin{equation}\label{F_s}
	\setlength{\abovedisplayskip}{2pt}
	\boldsymbol{f}_{\rm{s}}=\left[\boldsymbol{f}_{\rm{s}_1},...,\boldsymbol{f}_{{\rm{s}}_{n_i+1}} \right],
	\vspace{-0.5mm}
\end{equation}
where $\boldsymbol{f}_{\rm{s}_\textit{j}}$  is sparsely represented as ${\boldsymbol{W}_{\rm{s}_\textit{j}}}{\boldsymbol{e}_{\rm{s}_\textit{j}}}$, and $n_i+1$ is the number of the segments within road $\rm{r}_\textit{i}$. 
\subsection{Coordinate Mapping}
%曲线拟合，是坐标级的定位先验数据的建立，应加入实时更新以及抗噪声设计（噪声范围内映射）
To obtain a more fine-grained position of the vehicle within a segment, curve fitting is employed to map HetNet signals into two-dimensional coordinates. Specifically, $2K$ polynomial curve functions are utilized, where $K$ represents the number of HetNet BSs. The longitude curve function and the latitude curve function of the $k$th HetNet BS within the $l$th segment can be expressed as follows: 
\begin{equation}\label{cur}
	\vspace{-1mm}
	x_j^l = G_k^l \left(P_{j,k}^l\right) = \theta_{0,k}^l+\theta_{1,k}^l{P_{j,k}^l}+...+\theta_{w,k}^l\left(P_{j,k}^l\right)^w,
	\vspace{1mm}
\end{equation}
\begin{equation}\label{cury}
	\vspace{-1mm}
	y_j^l = H_k^l\left(P_{j,k}^l\right) = \alpha_{0,k}^l+\alpha_{1,k}^l{P_{j,k}^l}+...+\alpha_{w,k}^l\left(P_{j,k}^l\right)^w,
	\vspace{1mm}
\end{equation}
where $\theta_{w,k}^l$ and $\alpha_{w,k}^l$ $\left(w\in \left[1, W\right], k \in \left[1, K\right], l\in\left[1, n_i+1\right] \right)$ respectively denote the fitted longitude and latitude coefficients. In (\ref{cur}) and (\ref{cury}), $w$ represents the order of a polynomial, $k$ denotes the BS index, and $l$ is the segment index. In (\ref{cur}) (and (\ref{cury})), $x_j^l$ (and $y_j^l$) denotes the real longitude (and latitude) of the $j$th position on a road. $G_k^l \left(P_{j,k}^l\right)$ (and $H_k^l\left(P_{j,k}^l\right)$) is the estimated longitude (and latitude) based on the RSSs of the $k$th BS. The coefficients of (\ref{cur}) and (\ref{cury}) are determined by the least squares principle described in \cite{Wang2015Curve} to minimize fitting errors. The final longitude and latitude results are obtained by the average longitude and latitude values mapped by all HetNet RSS values, respectively.   
\section{Multi-scale Vehicle Localization}\label{Online}
To locate a vehicle dynamically, the vehicular SFs are first extracted from the real-time HetNet signal sequence using the sparse representation model. Then, by integrating inclusion and adjacency relationships between roads and segments, we propose the MSVL algorithm to locate the vehicular road, segment, and coordinate based on extracted SFs.
\subsection{Real-time Signal Feature Extraction}
To extract real-time SFs, the vehicular HetNet signal sequence of a segment is first acquired by SPs extracted in Section \ref{segprocess}. Then, the vehicular feature set vector $\boldsymbol{e}_{\rm{u}}$ can be extracted by the model in (\ref{featureset}) from the real-time HetNet signal sequence. The vehicular multi-scale SF vectors, i.e., $\hat{\boldsymbol{f}}_{\rm{r}_\textit{i}}$ and $\hat{\boldsymbol{f}}_{\rm{s}_\textit{l}}$, can be extracted by multiplying the vehicular feature set vector and the feature selection matrix of the matched road $\rm{r}_\textit{i}$ and segment $\rm{s}_\textit{l}$.
\subsection{Multi-scale Feature Matching}\label{Feature Matching}
To achieve high precision with low computation complexity, the posterior probability of a segment is utilized for vehicular localization by fusing geographical prior information. Based on the road feature vector $\boldsymbol{f}_{\rm{r}}$ and segment feature vector $\boldsymbol{f}_{\rm{s}}$, the vehicular posterior probability of a segment can be obtained by applying the Bayes’ rule as follows: 
%给出马尔可夫条件转移概率的总公式，然后再逐个解释每个概率的物理意义。

\begin{equation} \label{postpro}
	\begin{aligned}
	\setlength{\abovedisplayskip}{2pt}
	\hspace{-2mm}
	\rm{Pr}\left(\hat{\rm{s}_\textit{l}} \mid \rm{s}_\textit{l} \right) &= \rm{Pr}\left(\hat{\boldsymbol{f}}_{\rm{s}_\textit{l}} \mid \boldsymbol{f}_{\rm{s}_\textit{l}} \right) \\ &=\frac{\rm{Pr}\left(\hat{\boldsymbol{f}}_{\rm{r}_\textit{i}} \mid {\boldsymbol{f}_{\rm{r}_\textit{i}}}\right)
		\rm{Pr}\left(\rm{s}_\textit{l} \mid \rm{r}_\textit{i} \right)
		\rm{Pr}\left(\hat{\boldsymbol{f}}_{\rm{s}_\textit{l}} \right)}
	{\sum\limits_{q = 1}^{n} \rm{Pr}\left(\hat{\boldsymbol{f}}_{\rm{s}_\textit{q}} \mid \boldsymbol{f}_{\rm{s}_{\textit{q}}} \right)\rm{Pr}\left(\boldsymbol{f}_{\rm{s}_{\textit{q}}}\right)},
	\vspace{-1mm}
	\end{aligned}
\end{equation}
where $\rm{Pr}\left(\hat{\boldsymbol{f}}_{\rm{r}_\textit{i}} \mid {\boldsymbol{f}_{\rm{r}_\textit{i}}}\right)$ denotes the matching probability of road-scale SFs, $\rm{Pr}\left(\hat{\boldsymbol{f}}_{\rm{s}_\textit{l}} \mid \boldsymbol{f}_{\rm{s}_\textit{l}} \right)$ and $\rm{Pr}\left(\hat{\boldsymbol{f}}_{\rm{s}_\textit{q}} \mid \boldsymbol{f}_{\rm{s}_\textit{q}} \right)$ represent the matching probability of different segment-scale SFs. In the (\ref{postpro}),  $\rm{Pr}\left(\rm{s}_\textit{l} \mid \rm{r}_\textit{i} \right)$ denotes the prior probability of the segment $\rm{s}_\textit{l}$ within the road $\rm{r}_\textit{i}$. 
\par
Based on the posterior probability $\rm{Pr}\left(\hat{\rm{s}}_\textit{l} \mid \rm{s}_\textit{l} \right)$, the segment localization problem can be formulated as follows:
\begin{equation}\label{subsegprob}
	\setlength\abovedisplayskip{2pt}
	\setlength\belowdisplayskip{2pt}
	{\hat{\rm{s}}_\textit{l}} =\mathop{\arg\max}_{{\rm{s}_\textit{l}}  \in {\mathcal{S}_{i}}}
	\rm{Pr}\left(\hat{\rm{s}}_\textit{l} \mid \rm{s}_\textit{l} \right),
	%	\vspace{-1mm}
\end{equation}
To locate the vehicular positions with the highest posterior probability, we propose the MSVL algorithm which includes three modules, i.e., the road recognition module, the segment positioning module, and the coordinate localization module. 
\subsubsection{\textbf{Road recognition module}}
To recognize the real-time road SFs from the road-scale SF vector $\boldsymbol{f}_{\rm{r}}$, the vehicular SFs are matched with and each road SFs to maximize the road feature-matching probability. The road feature-matching probability  $\rm{Pr}\left(\hat{\boldsymbol{f}}_{\rm{r}_\textit{i}} \mid {\boldsymbol{f}_{\rm{r}_\textit{i}}}\right)$ fits to an exponential probability distribution \cite{newson2009hidden} that can be expressed as:
\begin{equation}
	\setlength{\abovedisplayskip}{3pt}
	\rm{Pr}\left(\hat{\boldsymbol{f}}_{\rm{r}_\textit{i}} \mid {\boldsymbol{f}_{\rm{r}_\textit{i}}}\right) = e^{-{d\left(\hat{\boldsymbol{f}}_{\rm{r}_\textit{i}}, \boldsymbol{f}_{\rm{r}_\textit{i}}\right)}},
	\vspace{-2mm}
\end{equation}
where $d\left(\hat{\boldsymbol{f}}_{\rm{r}_\textit{i}}, \boldsymbol{f}_{\rm{r}_\textit{i}}\right)$ represents the $l_2$-norm between the vector of vehicular SFs and the vector of $j$th road-scale SFs in $\boldsymbol{f}_{\rm{r}_\textit{i}}$. The $l_2$-norm $d\left({\hat{\boldsymbol{f}}_{\rm{r}_\textit{i}}}, \boldsymbol{f}_{\rm{r}_\textit{i}}\right)$ can be obtained as follows:
\begin{equation}
	\setlength{\abovedisplayskip}{4pt}
	d\left({\hat{\boldsymbol{f}}_{\rm{r}_\textit{i}}}, \boldsymbol{f}_{\rm{r}_\textit{i}}\right) = \|\boldsymbol{W}_{\rm{r}_\textit{i}}{\boldsymbol{e}_{\rm{u}}} - \boldsymbol{W}_{\rm{r}_\textit{i}}\boldsymbol{e}_{\rm{r}_\textit{i}}\|_2,
	\vspace{-1mm}
\end{equation}
where $\boldsymbol{W}_{\rm{r}_\textit{i}}\boldsymbol{e}_{\rm{r}_\textit{i}}$ denotes the $j$th road's SFs in the road-scale SF vector $\boldsymbol{f}_{\rm{r}}$, and $\boldsymbol{W}_{\rm{r}_\textit{i}}{\boldsymbol{e}_{\rm{u}}}$ represents the vehicular SFs corresponding to the $j$th road. With the highest probability of road-scale feature matching, real-time information about the vehicular road can be obtained.
\subsubsection{\textbf{Segment positioning module}}
To identify the specific location of the vehicular segment within the located road, the vehicular segment SFs are compared with the SFs in $\boldsymbol{f}_{\rm{s}}$ maximizing segment posterior probability. The probability of segment feature matching $\rm{Pr}\left(\hat{\boldsymbol{f}}_{\rm{s}_\textit{l}} \mid \boldsymbol{f}_{\rm{s}_\textit{l}} \right)$ can be given by:
\begin{equation}
	\setlength{\abovedisplayskip}{1pt}
	\rm{Pr}\left(\hat{\boldsymbol{f}}_{\rm{s}_\textit{l}} \mid \boldsymbol{f}_{\rm{s}_\textit{l}} \right) =  e^{-{d\left({\hat{\boldsymbol{f}}_{\rm{s}_\textit{l}}}, \boldsymbol{f}_{\rm{s}_\textit{l}}\right)}},
	\vspace{-2mm}
\end{equation}
where $d\left({\hat{\boldsymbol{f}}_{\rm{s}_\textit{l}}}, \boldsymbol{f}_{\rm{s}_\textit{l}}\right)$ represents the $l_2$-norm between the vector of vehicular segment SFs and the vector of the $l$th SFs in $\boldsymbol{f}_{\rm{s}}$. The $d\left({\hat{\boldsymbol{f}}_{\rm{s}_\textit{l}}}, \boldsymbol{f}_{\rm{s}_\textit{l}}\right)$ can be expressed as follows:
\begin{equation}\label{segdis}
	\setlength{\abovedisplayskip}{2pt}
	d\left({\hat{\boldsymbol{f}}_{\rm{s}_\textit{l}}}, \boldsymbol{f}_{\rm{s}_\textit{l}}\right) = \|\boldsymbol{W}_{\rm{s}_\textit{l}}\boldsymbol{e}_{\rm{u}} - \boldsymbol{W}_{\rm{s}_\textit{l}}\boldsymbol{e}_{\rm{s}_\textit{l}}\|_2,
	\vspace{-1mm}
\end{equation}
where $\boldsymbol{W}_{\rm{s}_\textit{l}}{\boldsymbol{e}_{\rm{u}}}$ represents the vehicular SFs. In (\ref{segdis}),  $\boldsymbol{W}_{\rm{s}_\textit{l}}\boldsymbol{e}_{\rm{s}_\textit{l}}$ denotes the SF vector of the $l$th segment in the $\boldsymbol{f}_{\rm{s}}$. By road and segment joint feature matching, the vehicular segment can be quickly located with the maximum segment posterior probability.
\subsubsection{\textbf{Coordinate localization module}} 
After acquiring the road and the segment by the road recognition module and the segment positioning module, the $j$th coordinate within the determined segment, i.e.,  $\hat{\boldsymbol{c}}_j=\left[\hat{x}_j^l,\hat{y}_j^l \right]^\top$ can be acquired as follows: 
\begin{equation}
	\vspace{-2mm}
	\hat{x}_j^l = \frac{1}{K} \sum\limits_{k = 1}^{K} G_k^l\left(P_{j,k}^l\right),
	\vspace{1mm}
\end{equation}
\begin{equation}
	\vspace{-1mm}	
	\hat{y}_j^l = \frac{1}{K} \sum\limits_{k = 1}^{K} H_k^l\left(P_{j,k}^l\right),
	\vspace{-1mm}
\end{equation}
where $\hat{x}_j^l$ and $\hat{y}_j^l$ denote the estimated longitude and latitude values. $G_k^l\left(P_{j,k}^l\right)$ and $H_k^l\left(P_{j,k}^l\right)$ represent the fitted curves obtained by (\ref{cur}) and (\ref{cury}) based on the $k$th BS signals. Subsequently, the multi-scale localization problem in (\ref{locproblem}) can be solved, and the real-time vehicular position $\boldsymbol{p} = \left[{\hat{\rm{r}}},{\hat{\rm{s}}_\textit{l}},{\hat{\boldsymbol{c}}_\textit{j}} \right]^\top$ can be localized with low latency and high precision.
\par
In the proposed MSVL algorithm, the highest computation complexity occurs when each road and segment is represented by all features of the feature set. The highest computation complexity is $\mathcal{O}\left(qK\left(m+n_i+1\right)\right)$, where $m$, $n_i+1$, and $qK$ represent the number of roads, the number of segments within the localized road, and the number of feature kinds in the feature set, respectively. As the number of feature kinds $q$ increases, the localization latency will correspondingly increase. This is because the computational cost of the proposed MSVL algorithm is linearly related to the number of features. With the increase in the feature number, the algorithm will need to process larger amounts of data, leading to an increase in computation time. Therefore, in practical applications, salient feature extraction is crucial to extract sparse features reducing computation complexity, and balancing mechanism positioning accuracy and positioning delay. 

\begin{algorithm}[!htp]
	\begin{spacing}{1}
		\small
		\caption{MSVL Algorithm}
		\label{alg5}
		\renewcommand{\algorithmicrequire}{\textbf{Input:}} \renewcommand{\algorithmicensure}{\textbf{Output:}}
		\begin{algorithmic}[1] 
			\Require The road-scale SF vector $\boldsymbol{f}_{\rm{r}}$, the segment-scale SF vector $\boldsymbol{f}_{\rm{s}}$, the real-time HetNet signal sequence $\left\{\boldsymbol{o}_1,\hdots,\boldsymbol{o}_t \right\}$, the vehicular feature set vector $\boldsymbol{e}_{\rm{u}}$.  
			\Ensure The estimated road $\hat{\rm{r}}_i$, segment $\hat{\rm{s}}_\textit{l}$, and the real-time coordinate $\hat{\boldsymbol{c}}$.
			\State   $\hat{\rm{r}}_i$ ← $\phi$, $\hat{\rm{s}}_\textit{l}$ ← $\phi$, $\hat{\boldsymbol{c}}=\left[\hat{x},\hat{y}\right]^\top$ ← $\phi$;		 
			\State The road matching probability set $\mathcal{D}_{\rm{r}_\textit{i}}$ ← $\phi$;
			\State The segment posterior probability set $\mathcal{D}_{\rm{s}_\textit{l}}$ ← $\phi$;			
			\For {$\boldsymbol{f}_{\rm{r}_\textit{i}}$ in $\boldsymbol{f}_{\rm{r}}$}  % For 语句，需要和EndFor对应
			\State Based on the Proposition \ref{CRLBfeatureset} and the Proposition \ref{featurereduce}, extract the vehicular road-scale SFs $\hat{\boldsymbol{f}}_{\rm{r}_\textit{i}}$;
			\State The road matching probability $\rm{Pr}\left(\hat{\boldsymbol{f}}_{\rm{r}_\textit{i}}\mid {r}_{i}\right)$;
			\State $\mathcal{D}_{\rm{r}_\textit{i}}$ ← $\mathcal{D}_{\rm{r}_\textit{i}} \cup \rm{Pr} \left(\hat{\boldsymbol{f}}_{\rm{r}_\textit{i}}\mid {\rm{r}_\textit{i}}\right)$; 			
			\EndFor
			\State The vehicular road $\hat{\rm{r}}_i$ with the maximum $\rm{Pr} \left(\hat{\boldsymbol{f}}_{\rm{r}_\textit{i}}\mid {r}_{i}\right)$; 	
			\For {$\boldsymbol{f}_{\rm{s}_\textit{l}}$ in $\boldsymbol{f}_{\rm{s}}$ of the road $\rm{r}_\textit{i}$}
			\State The vehicular segment-scale SFs $\hat{\boldsymbol{f}}_{\rm{s}_\textit{l}}$;							
			\State The segment posterior probability $\rm{Pr}\left(\rm{s}_\textit{l}\mid \hat{\boldsymbol{f}}_{\rm{s}_\textit{l}} \right)$;
			\State $\mathcal{D}_{\rm{s}_\textit{l}}$ ←${\mathcal{D}_{\rm{s}_\textit{l}}} \cup {Pr\left(\rm{s}_\textit{l}\mid \hat{\boldsymbol{f}}_{\rm{s}_\textit{l}} \right)}$;
			\EndFor			
			\State The vehicular segment $\hat{\rm{s}}_\textit{l}$ with the maximum ${Pr\left(\rm{s}_\textit{l} \mid \hat{\boldsymbol{f}}_{\rm{s}_\textit{l}} \right)}$;			 
			\State The vehicular coordinate $\hat{\boldsymbol{c}}=\left[\hat{x},\hat{y}\right]^\top$ within the positioned segment $\hat{\rm{s}}_\textit{l}$;
			\State \Return The vehicular road $\hat{\rm{r}}_i$, the segment $\hat{\rm{s}}_\textit{l}$, and the real-time coordinate $\hat{\boldsymbol{c}}$;
		\end{algorithmic} 
	\end{spacing}
\end{algorithm}
\par
The pseudo-code of the multi-scale vehicular localization process is presented in algorithm 1. The inputs of algorithm 1 are the multi-scale feature vector, a real-time vehicular RSS sequence, and the vehicular feature set vector. First, the vehicular road is located by SF matching between the vehicular road-scale SF vector and the extracted road-scale feature vector (steps 1-9). Then, the vehicular segment is located by maximizing the segment posterior probability based on SF matching (steps 10-15). Finally, the vehicular real-time coordinate within the located segment can be positioned through curve fitting (step 16). The outputs of algorithm 1 are the vehicular moving road, segment, and the real-time vehicular coordinate (step 17). 
\section{Experiment Results and Analysis}
\label{Experiment Results}
In this section, we conduct experiments to evaluate the localization performance of the proposed multi-scale localization mechanism, and the results are compared in terms of segment positioning error, coordinate positioning error, and coordinate mean localization latency.  
\subsection{ Simulation Setup}\label{Implementation}
The simulations are conducted in the Python 3.9 on a PC with 24 GB RAM. Unless otherwise stated, the system parameters are set as follows. The number of feature kinds $q$ in the proposed mechanism is set as $5$. Signals of $6$ HetNet BSs are used for vehicle localization, i.e., the maximum $K$ is equal to $6$. The distance interval of signal sampling $\Delta D$ is set as $1$ m. The sampling interval time $t$ and the vehicular speed $v$ are respectively set as $125$ ms and $30$ Km/h for signal collecting and real-time positioning simulation.
\par
In the conducted experiments, the proposed mechanism is evaluated within a HetNet utilizing the 4G frequency band ranging from 1800 MHz to 1900 MHz and the 5G frequency band spanning 3400 MHz to 3600 MHz.  Specifically, to accurately reflect the path loss in the 4G mobile systems, the Cost-231 Hata model \cite{7538693} is used to model the 4G channels based on the frequency band between 1800MHz and 1900MHz. This model can effectively incorporate urban architectural influences considering factors such as street widths, heights of surrounding buildings, and their spatial distribution, and allow for application across diverse environments, including urban, suburban, and rural settings. Meanwhile, the 1800MHz to 1900MHz frequency band in 4G has a higher signal transmission rate suitable for urban areas and densely populated places of vehicle positioning. Then, the 5G path loss model for our proposed mechanism is developed in alignment with TR 38.901 \cite{38901} operating on the frequency band between 3400MHz to 3600MHz, where various channel models tailored for diverse scenarios, including train, vehicle, urban, and rural environments can be outlined.
 
\par
Four approaches are implemented for performance comparison, i.e., the state of art restricted weighted K-nearest neighbor algorithm (RWKNN) \cite{2022RWKNN}, the hierarchical localization method (HLM) \cite{Hierarchical}, the gradient-based fingerprinting (GIFT) \cite{Gradient}, and the curve fitting-based exhaustive location search algorithm (CF-ELS) \cite{Wang2015Curve}. The former two methods are compared since they respectively utilize uniform grid signals and hierarchical signals to locate vehicles. The GIFT is chosen due to VUE can be located through gradient feature matching. The CF-ELS is chosen to be compared as range estimations are utilized in VUE localization. The number of nearest neighbors is set as $3$ in RWKNN \cite{2022RWKNN}. Two hidden layer sizes of the auto-encoder applied in HLM are set as 100 and 6, respectively \cite{Hierarchical}. The step size of the exhaustive search in CF-ELS is set to $0.1$ m \cite{Wang2015Curve}.
\par
\begin{figure}[t]
	\centering
	\setlength{\abovecaptionskip}{0.cm}
	\includegraphics[width=7.1cm]{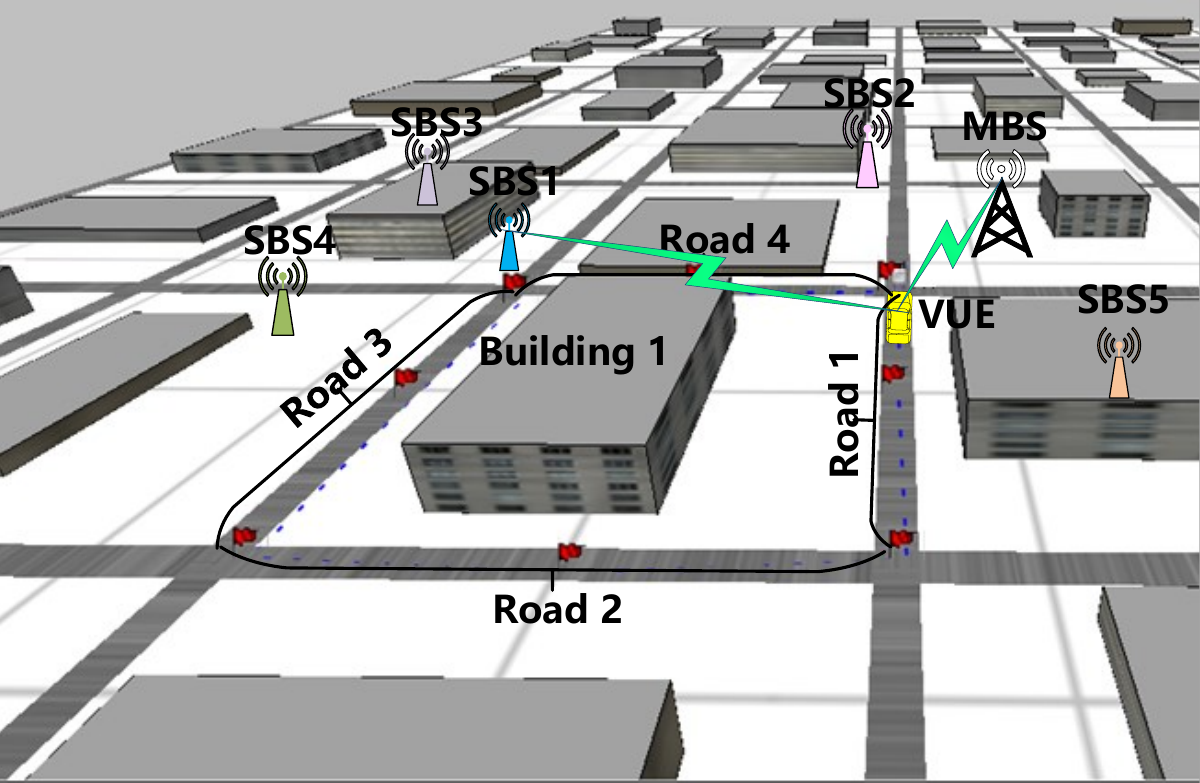}
	\vspace{2mm}
	\caption{Localization scenario in the Qualnet simulator.}
	\vspace{-2mm}
	\label{figure4}
\end{figure}
\begin{figure}[t]
	\centering
	\setlength{\abovecaptionskip}{0.cm}
	\includegraphics[width=7.1cm]{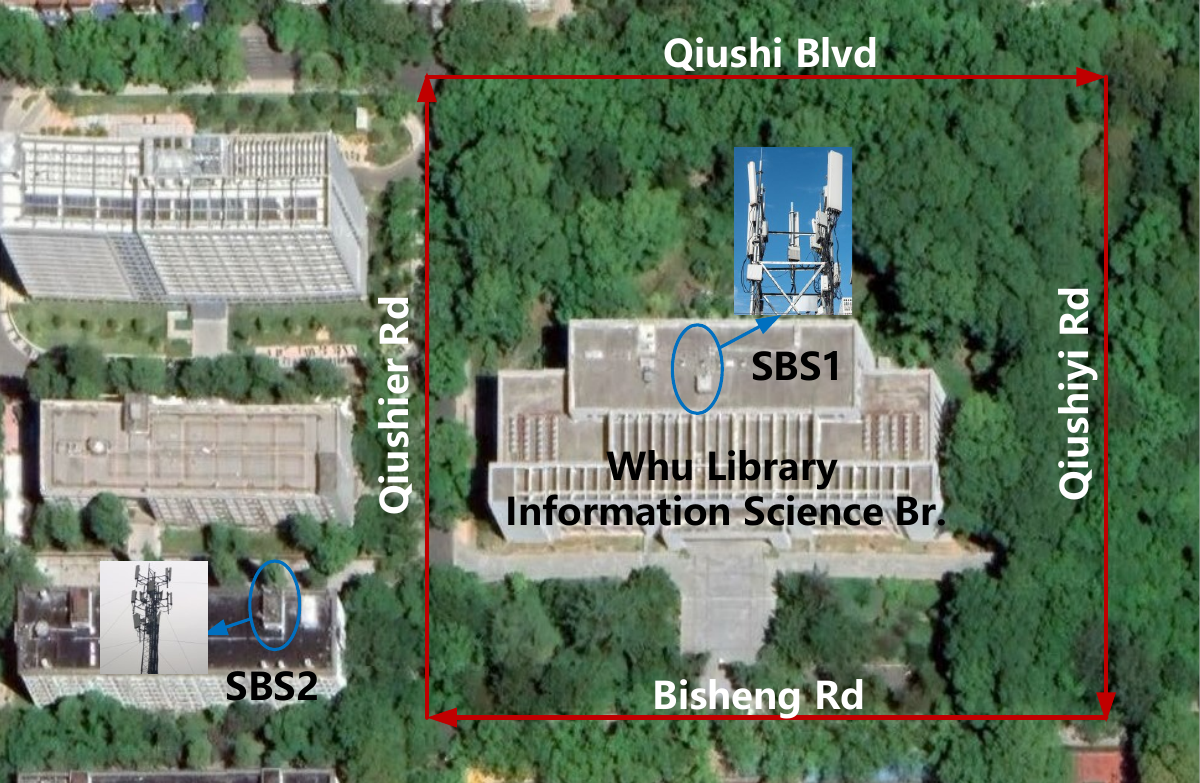}
	\vspace{2mm}
	\caption{Localization scenario in an urban environment.}
	\vspace{-4mm}
	\label{figure14}
\end{figure}

%为什么采用两个定位场景进行实验验证
To evaluate that the positioning performance in wide localization environments, an interested area of 0.6 Km $\times$ 0.6 Km was established by the Qualnet simulator based on real service parameters of HetNet BSs \cite{RECOMMENDATIONIP}, as shown in Fig. \ref{figure4}. There are $1$ MBS and $5$ SBS in the localization scenario. Building 1 is an urban high-rise building and its height is $65$ m. The width of each road in Fig. \ref{figure4} is $20$ m. To further verify the positioning performance of the road-aware positioning mechanism in real road environments, a real-world area of 0.15 Km $\times$ 0.15 Km was selected at Wuhan University, as shown in Fig.\ref{figure14}. The vehicular positions are estimated when the vehicle moves around building 1 along four roads as shown in Fig. \ref{figure4}.  In the real-world urban road environment, the vehicle will be located while moving on the four roads as shown in Fig.\ref{figure14}. 
\par
To assess the effectiveness of the proposed mechanism and the benchmark methods in terms of localization, both simulated and real experiments are conducted by extracting the simulated data set and the real-world data set, respectively. Specifically, the simulation data set was collected by signal measurement reports in the Qualnet simulator, with 13932 records obtained from the targeted localization area. In the simulation data set, each record contains RSS values from 6 HetNet BSs along with a 2D coordinate. In the real experiments, we collect the real-world data set along different roads through the Great Scott Gadget HackRF \cite{Giovanni2018Passive}, which can be able to capture radio signals from 1 MHz to 6 GHz and whose signal sensitivity can reach as low as -120 dBm. This data set contains 6495 records, where each record contains a two-dimensional coordinate and RSS values from 2 HetNet BSs. In the real-time vehicle positioning stage, the VUE received signals can be extracted by an RF measurement tool (Cellular-Z) \cite{s20061691}.
\subsection{Segment Positioning Performance} 
\begin{figure}[t]
	\centering
	\setlength{\abovecaptionskip}{0.cm}
	\includegraphics[width=8 cm]{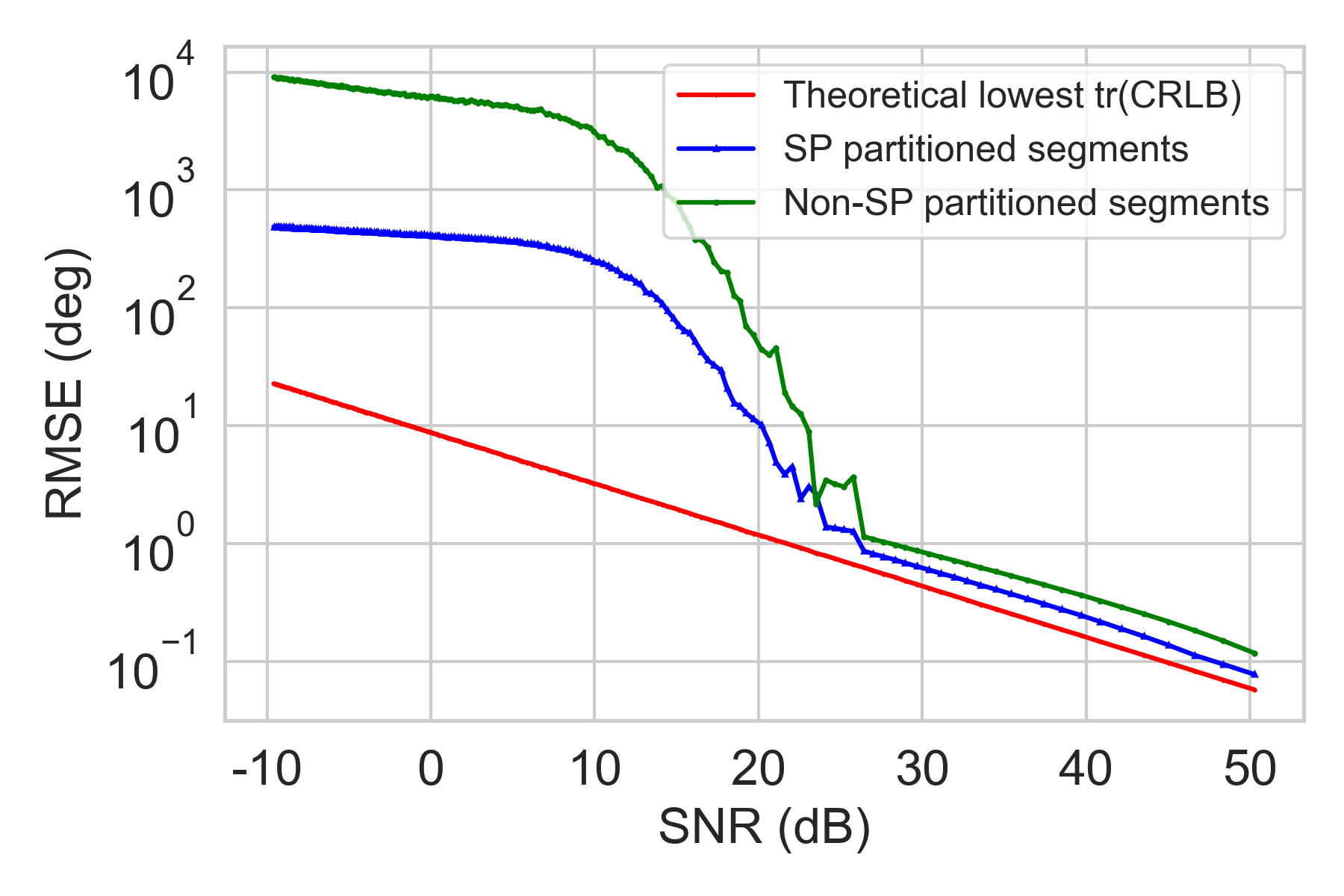}
	\vspace{1mm}
	\caption{RMSE of different partitioning segments.}
	\vspace{-3mm}
	\label{figure11}
\end{figure}
In Fig. \ref{figure11}, the root mean squared error (RMSE) is simulated with the increasing SNR of the service BS to evaluate the effectiveness of the proposed SP segmentation method. The SNR is defined as SNR = $10 \log \frac{\sigma^2_{Signal}}{\sigma^2_{Noise}}$, where $\sigma^2_{Signal}$ and $\sigma^2_{Noise}$ respectively are the signal variance and the noise variance. The SP partitioned segments are obtained using the proposed SP segmentation method based on the Proposition \ref{CRLBsegpo} and Remark \ref{segbeta}. The non-SP partitioned segments are divided by the signals are not the extreme points and turning points of a HetNet signal sequence. In Fig. \ref{figure11}, the RMSE of the segments positioned by SPs can approach the lowest CRLB as the SNR increases, indicating that SP partitioned segments are more resistant to noise during the positioning process. The result in Fig. \ref{figure11} also demonstrates that the proposed sequence segmentation method can accurately partition segments and indeed reduce the segment localization error.  

\begin{figure}[t]
	\centering
	\setlength{\abovecaptionskip}{0.cm}
	\includegraphics[width=8 cm]{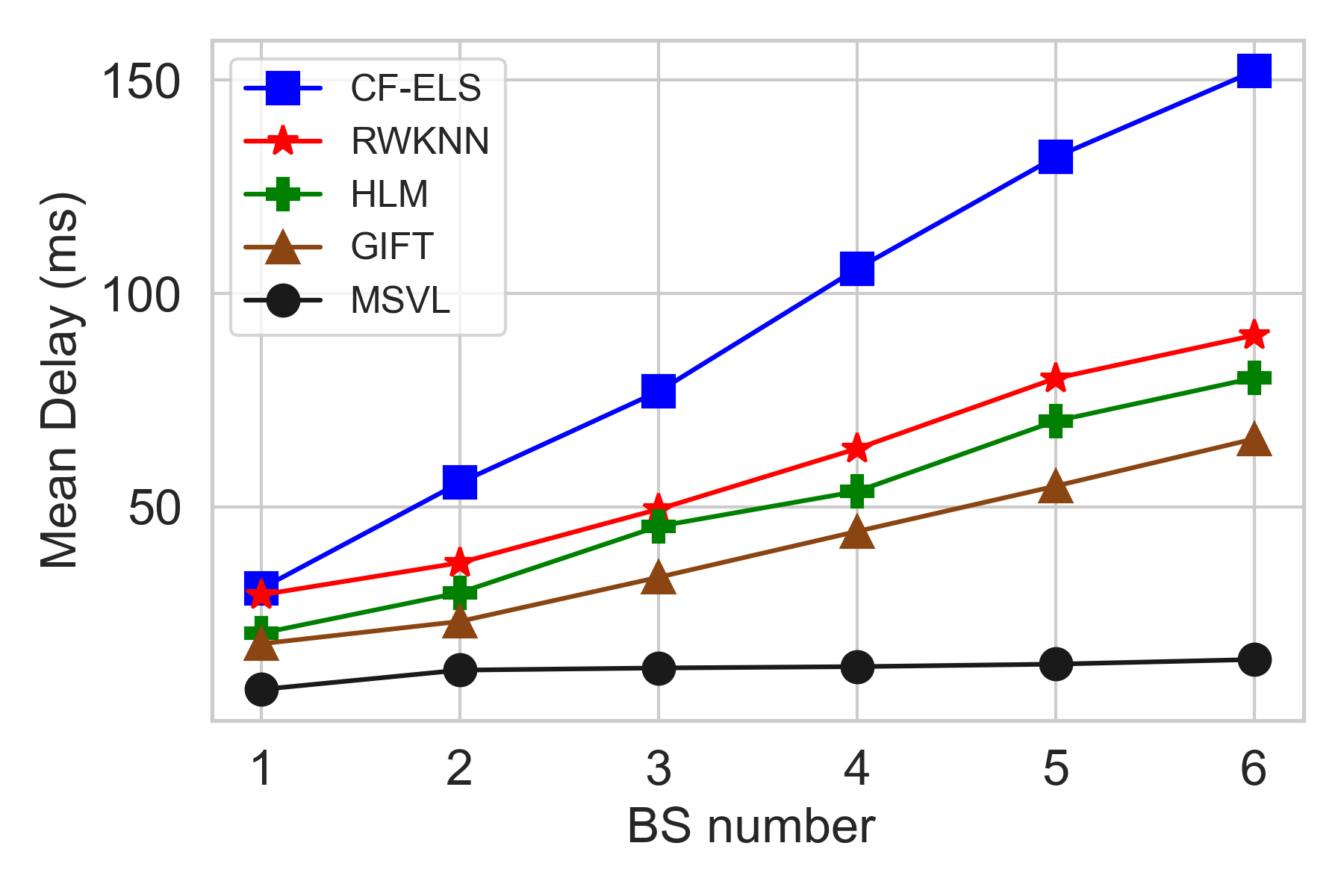}
	\vspace{1mm}
%	\captionsetup{labelfont={color=red}}
	\caption{Mean delay versus BS number.}
	\vspace{-4mm}
	\label{figure6}
\end{figure} 
\begin{figure}[t]
	\centering
	\setlength{\abovecaptionskip}{0.cm}
	\includegraphics[width=8 cm]{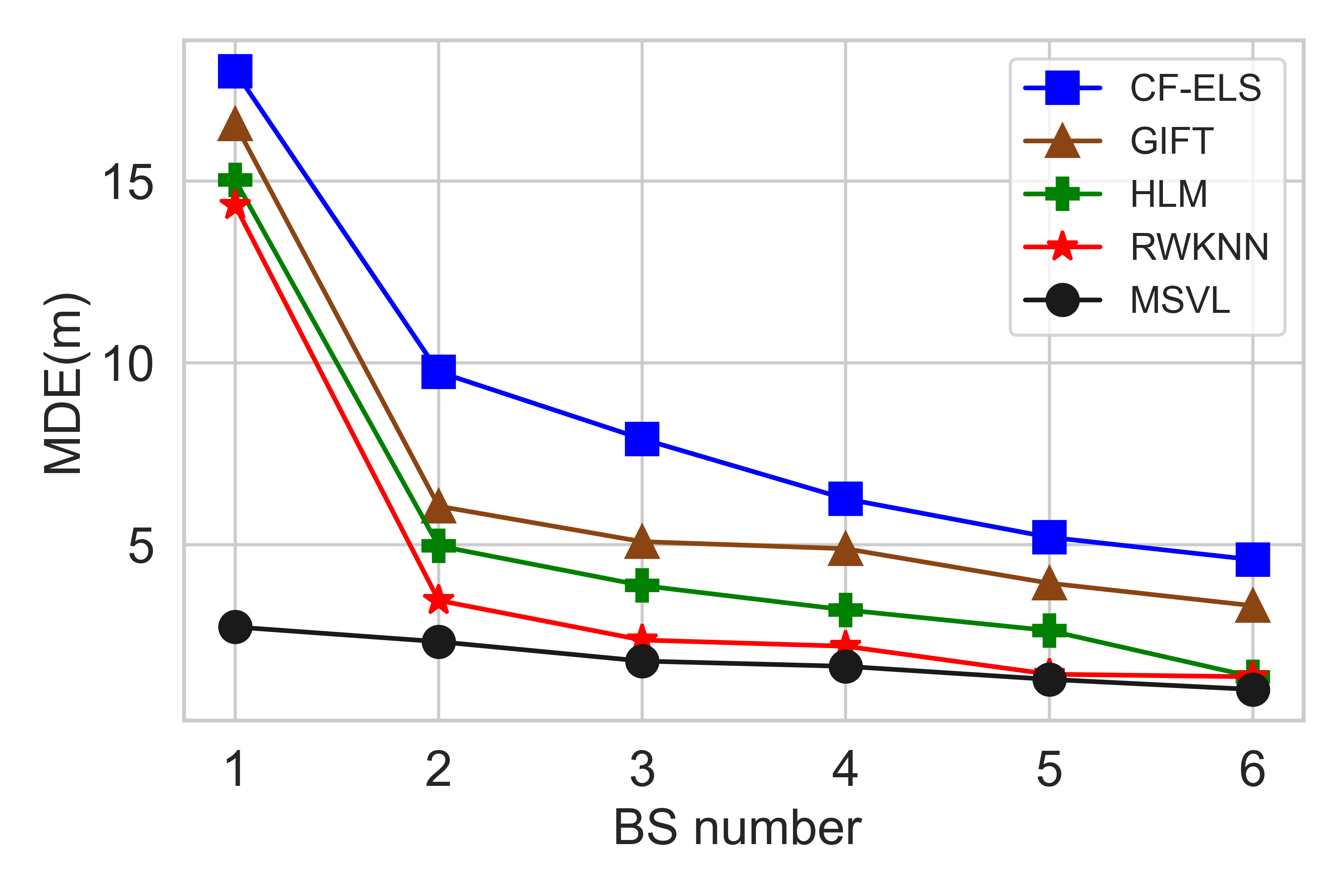}
	\vspace{1mm}
%	\captionsetup{labelfont={color=red}}
	\caption{Mean error versus BS number.}
	\vspace{-2mm}
	\label{figure5}
\end{figure}

\begin{figure}[t]
	\centering
	\setlength{\abovecaptionskip}{0.cm}
	\includegraphics[width=8 cm]{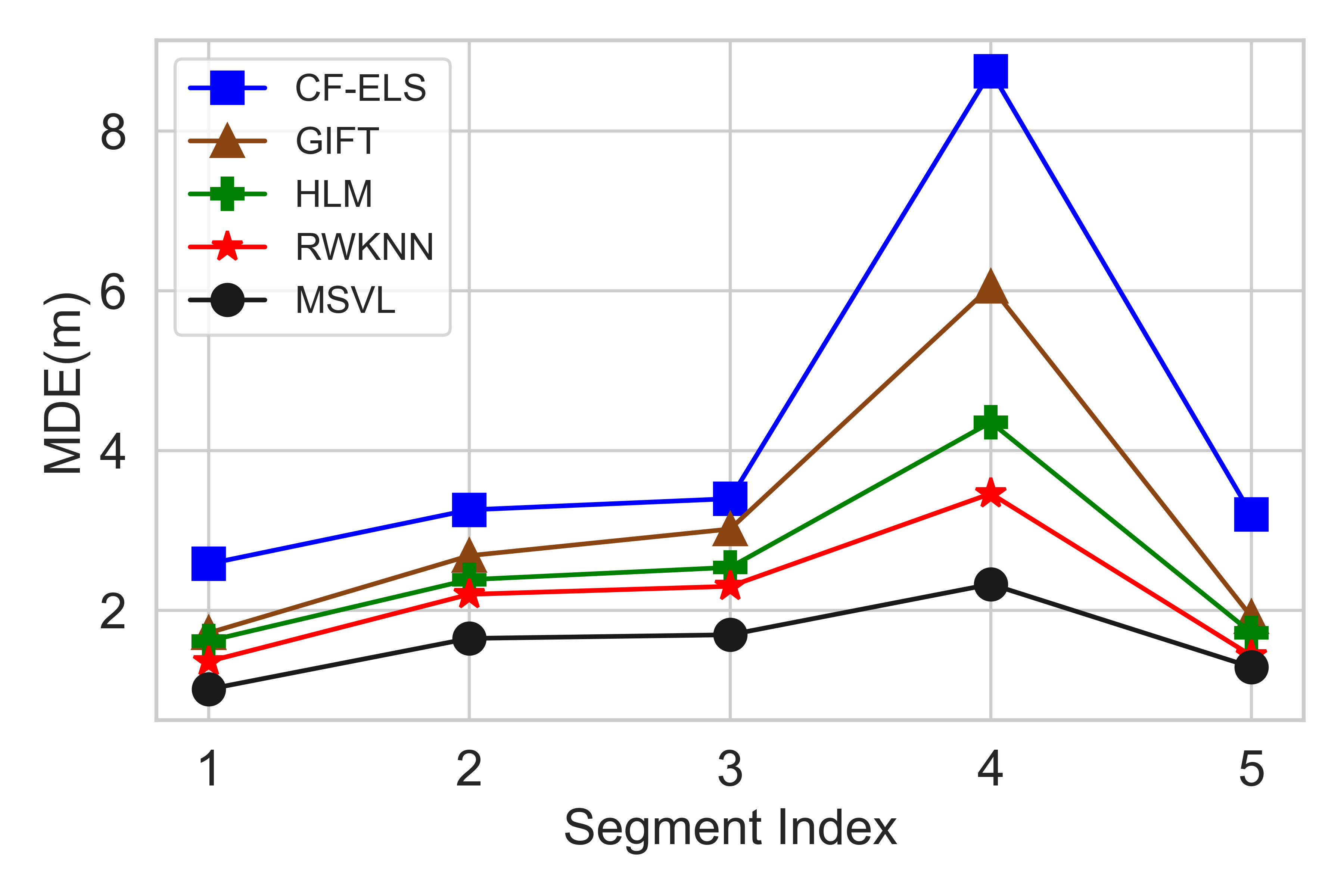}
	\vspace{1mm}
%	\captionsetup{labelfont={color=red}}
	\caption{MDE of different segments with varying BS number.}
	\vspace{-2mm}
	\label{figure12}
\end{figure}

\vspace{2mm}
\begin{figure}[t]
	\centering
	\setlength{\abovecaptionskip}{0.cm}
	\includegraphics[width=8 cm]{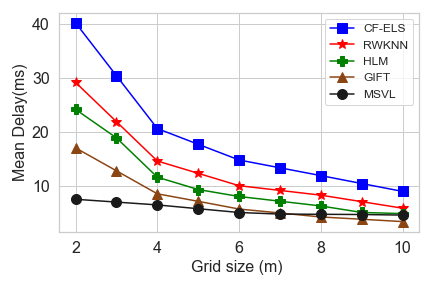}
	\vspace{1mm}
%	\captionsetup{labelfont={color=red}}
	\caption{Mean delay versus grid size.}
	\vspace{-2mm}
	\label{figure8}
\end{figure}
\begin{figure}[t]
	\centering
	\setlength{\abovecaptionskip}{0.cm}
	\includegraphics[width=8 cm]{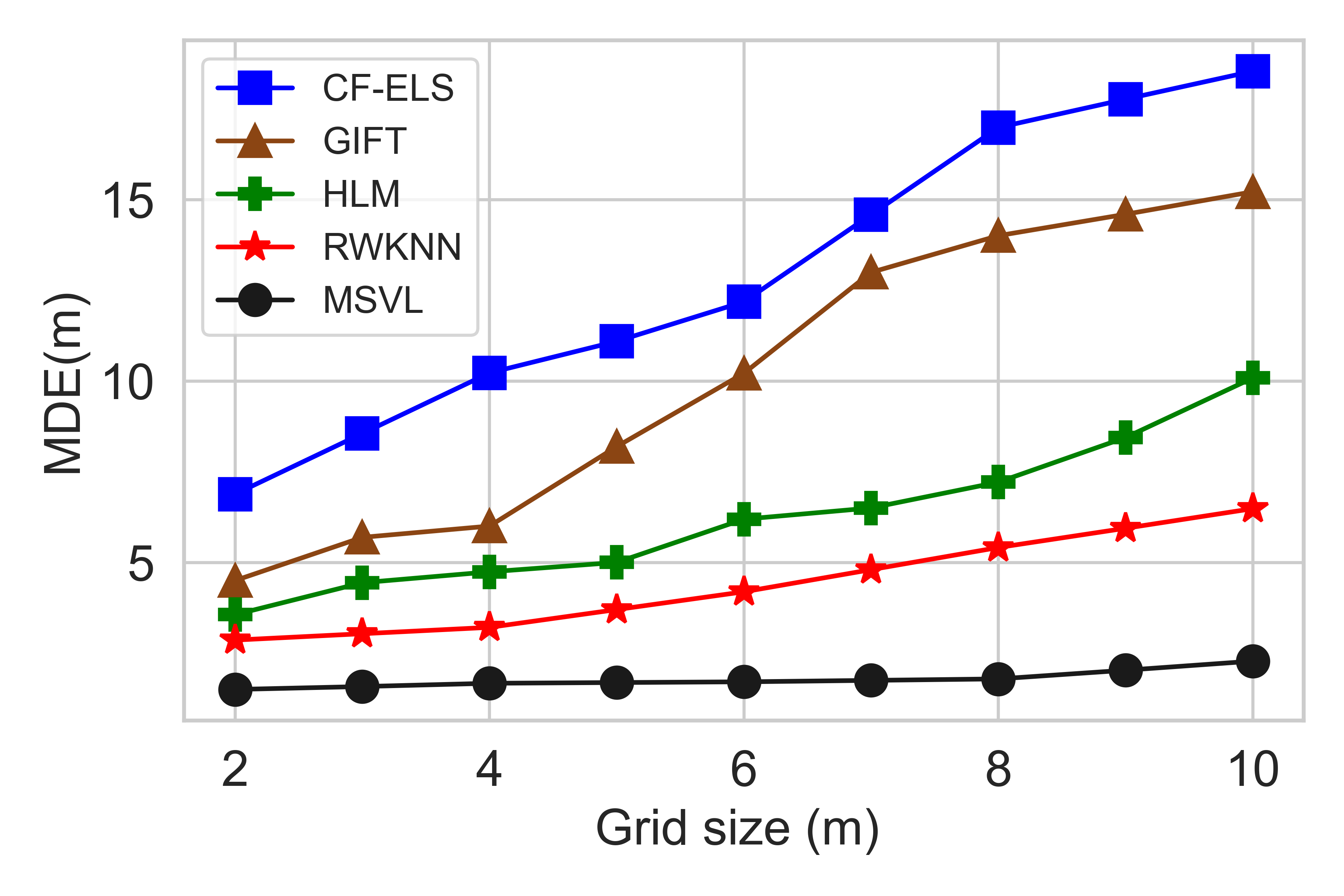}
	\vspace{1mm}
%	\captionsetup{labelfont={color=red}}
	\caption{Mean error versus grid size.}
	\vspace{-4mm}
	\label{figure7}
\end{figure}

\begin{figure}[t]
	\centering
	\setlength{\abovecaptionskip}{0.cm}
	\includegraphics[width=8 cm]{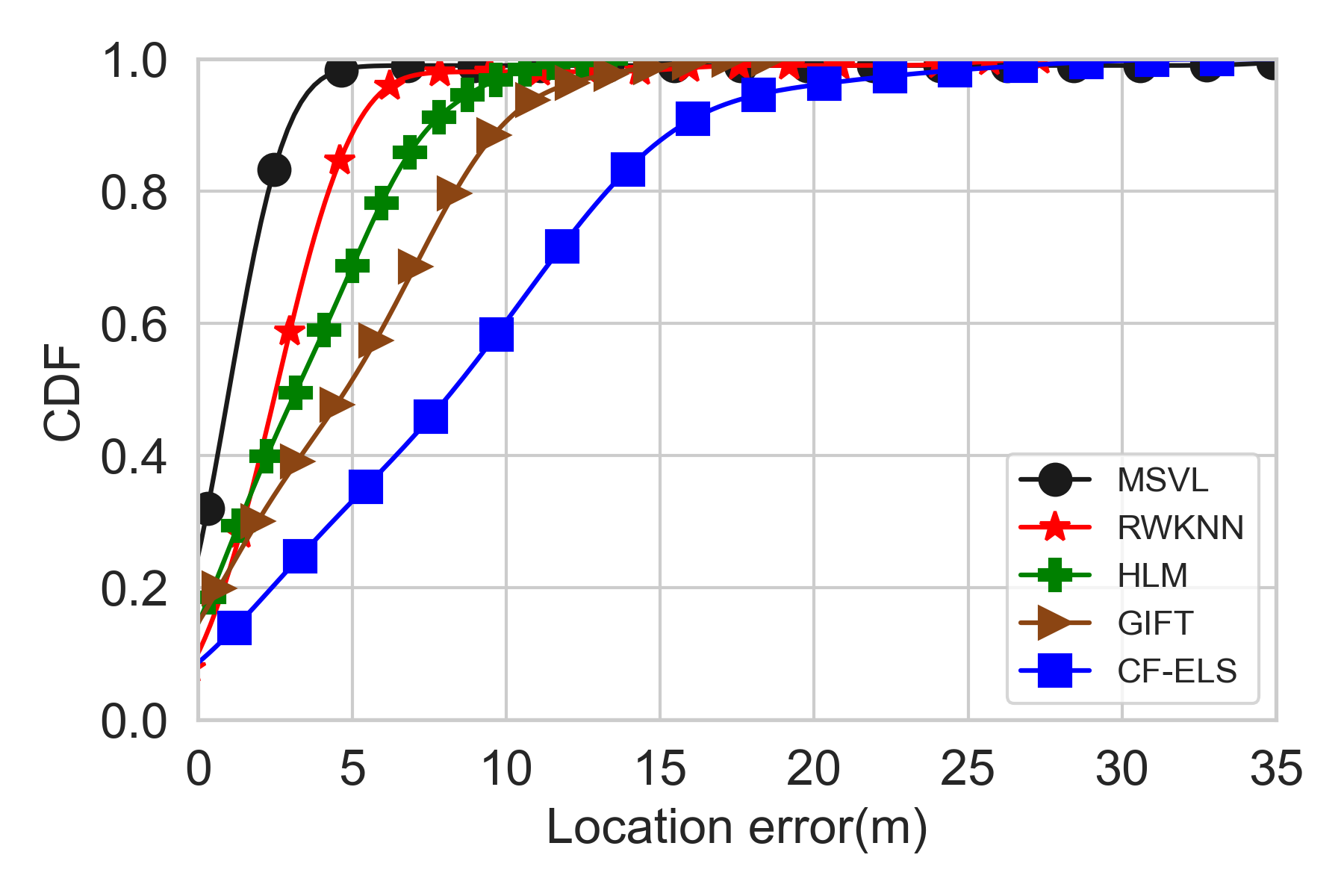}
	\vspace{-1mm}
%	\captionsetup{labelfont={color=red}}
	\caption{CDF based on simulated signal measurements.}
	\vspace{-3mm}
	\label{figure10}
\end{figure}
\vspace{-3mm}
\subsection{Coordinate Positioning Performance} 
Since the detectable BS number and the grid size play crucial roles in accurate signal feature extraction and computation complexity, simulation experiments were conducted to investigate the impact of these two parameters on the mean delay and the mean distance error (MDE). The mean delay is calculated by $\overline{T} = \frac{1}{N_l}\sum\limits_{i = 1}^{N_l} {T_i}$, and the MDE is defined as MDE $= \frac{1}{N_l}\sum\limits_{i = 1}^{N_l} {\Delta E_i}$, where $T_i$ and $\Delta E_i$ respectively denote the localization latency and the localization error of the $i$th position. $N_l$ represents the number of estimated target positions within the $l$th segment. 
\subsubsection{Impact of the HetNet BS number}
In Figs. \ref{figure6} and \ref{figure5}, the vehicular mean delay and the MDE of coordinate positions are illustrated respectively under different HetNet BS numbers at the grid size of 2 m for the proposed multi-scale localization mechanism and benchmark schemes. Specifically, it can be observed from Fig. \ref{figure6} that the proposed mechanism achieves the lowest mean delay compared with the benchmark method. This is due to the significantly reduced computation complexity, achieved by the utilization of low-dimension road and segment positioning space, as well as the sparse SFs in the feature-matching localization. The effect of the different BS numbers on the MDE of the proposed mechanism outperforms benchmark methods is shown in Fig. \ref{figure5}. It is observed from Fig. \ref{figure5} that the proposed mechanism can achieve a minimum MDE of $2.43$ m when VUE receives signals from 2 BSs, owing to the substantial signal attenuation in the 5G HetNet. Since the RWKNN, HLM, GIFT, and CF-ELS methods are respectively based on uniform grids, dual-layer fingerprints, single gradient features, and triangulation positioning, these methods exhibit higher latency or positioning errors under different BS numbers.
\par
To evaluate the localization performance of our proposed mechanism in large environments, the experiment of localization error under different BS numbers is conducted. As shown in Fig. \ref{figure12}, a vehicle was moving on a road from the segment 1 to the segment 5, which is represented by segment indexes. The detected BS numbers of the segment 1 to the segment 5 are 6, 4, 4, 2, and 5, respectively. It can be seen that when the vehicle moves on the segment 4 which can only detect a MBS and a SBS, the MDEs of the benchmark methods are larger than the MDE of the proposed mechanism. This is because we mainly utilize the different properties of HetNet BSs rather than rely on absolute signal values.
\subsubsection{Impact of Different Grid Sizes}
In Figs. \ref{figure8} and \ref{figure7}, we show the mean delay and the MDE of coordinate positions under different grid sizes for the proposed multi-scale localization mechanism and benchmark schemes, respectively. Fig. \ref{figure8} shows that the mean delay decreases as the grid size increases. Specifically, the lowest mean delay is achieved by the proposed mechanism at the smallest grid size of 2 m, measuring less than $10$ ms. The reason is that roads and segments are divided according to the actual length of the road and HetNet signal features, thereby avoiding reliance on uniform grid sizes. Meanwhile, MDE of the proposed algorithm gradually decreases as the grid size increases and remains below $2.5$ m, as depicted in Fig. \ref{figure7}. Since the RWKNN, HLM, GIFT, and CF-ELS methods rely on uniform grids for signal feature extraction and vehicle localization, the MDEs of benchmark methods would increase as the grid size increases. In contrast, the proposed mechanism can achieve low-latency and high-precision vehicle localization under large grid sizes. 

\par
Based the above simulations of grid size on positioning latency and accuracy, the following analysis is provided regarding the sampling interval and vehicle speed that determine the grid size. To obtain a minimum grid of 1 m for evaluating positioning performance, the signal sampling time interval and vehicle speed were set to $125$ ms and $30$ Km/h, respectively. Under the constant sampling time interval $t$, increasing vehicle speed $v$ may reduce localization accuracy in signal collection and real-time positioning processes. The reason is that higher speeds accentuate the Doppler effect, resulting in carrier frequency offsets and rapid fading of received wireless signals, which distorts signal features. Moreover, larger sampling intervals may result in the loss of critical information, further diminishing localization precision. Doppler compensation methods \cite{8359368} and lower signal sampling time intervals can be used to solve the above issues for maintaining accurate positioning. When the vehicle speed remains constant, an increase in the signal sampling time interval results in a larger sampled distance interval. This leads to incorrectly partitioning of signal segments and inaccuracies in segment-scale features, ultimately diminishing the positioning accuracy of the proposed mechanism. Therefore, a small signal sampling time interval is utilized for maintaining precise signal representation and overall mechanism performance. 
\par
Fig. \ref{figure10} shows the cumulative distribution function (CDF) of localization errors using 2 HetNet BSs at the $2$ m grid size. The proposed multi-scale vehicle localization algorithm achieves improvements of $25\%$, $33\%$, $45\%$, and $77\%$ (from 2.6 m to 0 m) compared to RWKNN, HLM, GIFT, and CF-ELS in terms of distance estimation error.
\par
To intuitively reflect the real-time positioning process of vehicles through our mechanism in HetNet, we have made the experimental process  into a short movie \cite{movie}. In this movie, a vehicle is positioned by our proposed mechanism in a crowed urban road environment. The real-time HetNet signals and their associated features are first extracted. Then, the proposed MSVL algorithm is employed to determine roads, segments, and coordinates, ensuring accurate navigation within complex urban road environments. 
\subsubsection{Performance Comparison under the Real Data Set}
\begin{figure}[t]
	\centering
	\includegraphics[width=8 cm]{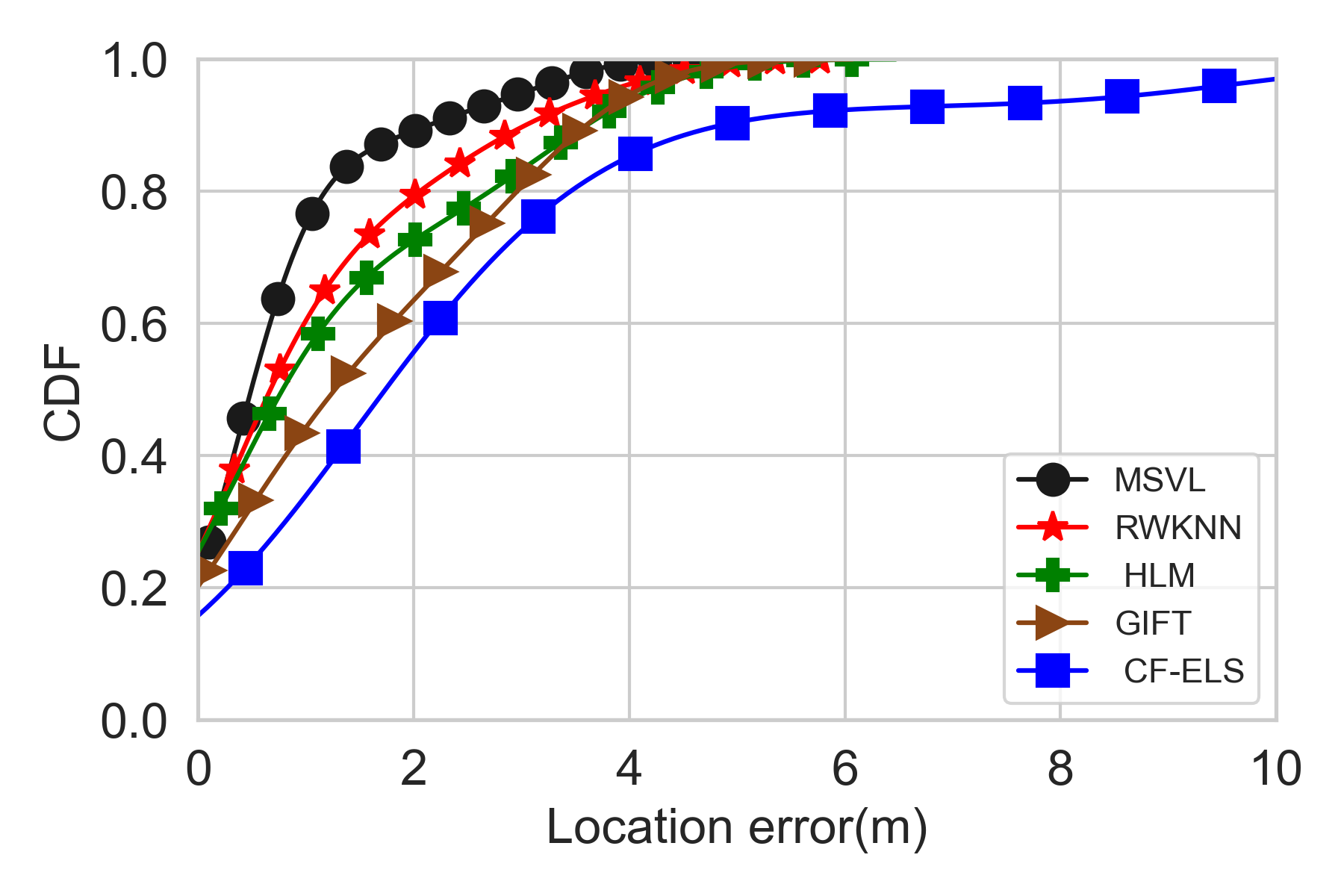}
	\vspace{-2mm}
%	\captionsetup{labelfont={color=red}}
	\caption{CDF based on real signal measurements.}
	\label{figure9}
	\vspace{-3mm}
\end{figure}

To conduct real experiments on vehicle positioning in urban road environments, the CDF of localization errors are evaluated based on the real-world data set collected from a HetNet scenario. It can be seen from Fig. \ref{figure9} that our proposed algorithm shows a $17\%$, $21\%$, $27\%$, and $35\%$ improvement (from 2 m to 0 m) respectively compared to RWKNN, HLM, GIFT, and CF-ELS.
\par
Table \ref{locationtime} depicts the mean delay of the four localization methods using the real-world data set. As the searching space is divided into the road searching space and the segment searching space, our proposed method's mean delay could be reduced to $7.2$ ms. For high-speed vehicles, a localization delay under $10$ ms is crucial for maintaining a safe distance between vehicles \cite{SelfDriving}. Given that the localization times of RWKNN, HLM, GIFT, and CF-ELS exceed $10$ ms, meeting the vehicle localization requirements with low latency and high reliability becomes challenging.
\begin{table}[h]	
	\centering	
	\vspace{-1mm}
%	\captionsetup{labelfont={color=red}}
	\caption{Mean delay of located positions}
	\vspace{1mm}
	\begin{tabular}{|c|c|c|c|c|c| }
		\hline 
		Method & {RWKNN} & HLM & GIFT & \tabincell{c}{CF-\\ELS} & {MSVL}\\
		\hline 
		\tabincell{c} {Mean \\ delay (ms)} & 32.7 & 30.1 & 28.5 & 57.8 & 7.2 \\ 
		\hline	 
	\end{tabular}
	\vspace{-1mm}
	\label{locationtime}
\end{table}
\par
In the above simulation experiments, the localization performance of the MSVL algorithm can achieve low latency and high precision in single-lane environments. In multi-lane road environments, localization errors of our proposed mechanism can be reduced by increasing BS density and road infrastructures. Specifically, through increasing BS density, more BS signals with differences can be utilized to locate vehicles, thereby reducing localization errors. In the field of intelligent transportation, to achieve low-latency and high-reliability vehicle positioning, more road side communicating and sensing infrastructures can be constructed for lane-level positioning and automatic driving.

\section{Conclusions}
\setlength\abovedisplayskip{1pt}
\setlength\belowdisplayskip{1pt}
\label{Conclusions}
In this paper, we proposed a multi-scale localization mechanism achieving low-latency vehicle localization. Specifically, a SP segmentation method and a SF extraction method were devised to sparsely represent roads and segments. Then, a MSVL algorithm was proposed through road and segment SF matching. Fine-grained positioning was achieved through curve fitting to localize vehicular coordinates. The numerical results confirmed that our proposed mechanism offers a promising solution to reduce localization latency and support reliable localization accuracy. The impact of faint HetNet signal variations on the localization performance of the proposed mechanism is a potential concern. However, this issue can be addressed by utilizing the different signals from more HetNet BSs. As part of future work, with the help of high precision beam forming, accurate lane-level positioning can be achieved by aiming the BS beam at the lane and then extracting lane-level signal features.
%\vspace{-7mm}
\section*{Appendix A: \textsc{Detailed Calculation of CRLB}}
\label{ProPropo1}
To locate the $l$th segment, suppose distance interval of signal sampling is 1 m, i.e., $\Delta D = \sqrt{ \left(x_{k+1}^l-x_k^l\right)^2+\left(y_{k+1}^l-y_k^l\right)^2 } = 1$ m, the mean square gradient feature of the $k$th BS RSSs within the $l$th segment in (\ref{plm}) can be expressed as follows: 
\begin{equation}\label{gradientfeature}
	\vspace{-1mm}
	\overline{{g_k^l}^{_2}}  = \frac{(10 \beta_k^l)^2}{N_l-1}\sum\limits_{i = 1}^{N_l-1} {\left( \log d_{i+1 ,k}^l-\log d_{i, k}^l \right)}^2 + Z_k^l,		
	\vspace{1mm}
\end{equation}
where $d_{i, k}^l$ is the distance between the $i$th position in the segment and the position of the $k$th BS. $d_{i, k}^l$ is calculated by the Euclidean distance as follows:
\begin{equation}\label{distance}
%	\vspace{-1mm}
	d_{i, k}^l  = \sqrt{ (x_i^l-x_{BS_k})^2+(y_i^l-y_{BS_k})^2 },
%	\vspace{-1mm}
\end{equation}
where $x_i^l$ and $y_i^l$ respectively represent the longitude value and the latitude value of the $i$th position within the $l$th segment. $x_{BS_k}$ and $y_{BS_k}$ respectively represent the longitude value and the latitude value of the $j$th BS. To find the coordinate derivative of segment positioning, coordinates within the $l$th segment are mapped to the middle position as follows:
\begin{equation}\label{mid}
	\vspace{-1mm}
	x_i^l = \frac{2i}{N_l}(x_m^l-x_1^l)+x_1^l, 
	\vspace{1mm}
\end{equation}
\begin{equation}\label{mid2}
	\vspace{-1mm}
	y_i^l = \frac{2i}{N_l}(y_m^l-y_1^l)+y_1^l, 
	\vspace{-1mm}
\end{equation}
where $x_m^l$ and $y_m^l$ respectively represent the longitude value and latitude value of the middle position in the $l$th segment. And $x_1^l$ and $y_1^l$ are the 2D coordinate of the first position in the $l$th segment. The Jacobian matrice of all BSs measurement errors evaluated at the true segment $\boldsymbol{c}_m^l = \left[ x_m^l, y_m^l \right]^\top$ takes the form
\begin{equation}\label{jacobi}
	\vspace{1mm}
	{J_{\overline{g^{_2}}}} = 
	\begin{pmatrix}
	\frac{\partial \overline{{g_{1}^l}^{_2}} }{\partial {x_m^l}} & \frac{\partial \overline{{g_{1}^l}^{_2}} }{\partial {y_m^l}} \\
	\vdots & \vdots \\
	\frac{\partial \overline{{g_{K}^l}^{_2}} }{\partial {x_m^l}}&
	\frac{\partial \overline{{g_{K}^l}^{_2}} }{\partial {y_m^l}}	
	\end{pmatrix}. 
%	\vspace{-1mm}
\end{equation}
The derivative of the mean square gradient feature can be expressed by 
\begin{equation}\label{deriv1}
	\vspace{-1mm}
	\begin{split}
		\vspace{-1mm}
		&\frac{\partial \overline{{g_{k}^l}^{_2}} }{\partial {x_m^l}} = 
		\frac{(10\beta_k^l)^2}{(N_l-1)\ln 10}[\sum\limits_{i = 1}^{N_l-1} {2\left( \log {d_{i+1,k}^l}-\log {d_{i,k}^l} \right)} \\&
		\left( \frac{\cos{\theta_{i+1,k}^l}}{d_{i+1,k}^l}
		\frac{2(i+1)}{N_l}-\frac{\cos{\theta_{i,k}^l}}{d_{i,k}^l}
		\frac{2i}{N_l} \right)],
	\end{split}
	%	\vspace{-1mm}
\end{equation}
\begin{equation}\label{deriv2}
	\vspace{-1mm}
	\begin{split}
		\vspace{-1mm}
		&\frac{\partial \overline{{g_{k}^l}^{_2}} }{\partial {y_m^l}} = 
		\frac{(10\beta_k^l)^2}{(N_l-1)\ln 10}[\sum\limits_{i = 1}^{N_l-1} {2\left( \log {d_{i+1,k}^l}-\log {d_{i,k}^l} \right)} \\&
		\left( \frac{\sin{\theta_{i+1,k}^l}}{d_{i+1,k}^l}
		\frac{2(i+1)}{N_l}-\frac{\sin{\theta_{i,k}^l}}{d_{i,k}^l}
		\frac{2i}{N_l} \right)],
	\end{split}
	%	\vspace{-1mm}
\end{equation}
where $\theta_{i,k}^l$ is the azimuth angle between the $i$th position and the $k$th BS within the $l$th segment.
The noise covariance matrix of the mean square gradient feature is
\begin{equation}\label{noise1}
	\Sigma_{\overline{{g}^{_2}}} = diag [ {\rho_1^l}^2,\cdots,{\rho_K^l}^2]_{N_l\times N_l}. 
\end{equation}
The Fisher information matrix (FIM) and the corresponding CRLB of all BSs' measurements take the form
\begin{equation}\label{c1}
	\Phi = J_{\overline{g^{_2}}}^\top \Sigma_{\overline{g^{_2}}}^{-1} J_{\overline{g^{_2}}},
\end{equation}
\begin{equation}\label{c2}
	{\rm{CRLB}} = \Phi^{-1}.
\vspace{-2mm}	
\end{equation} 
Since the tr(CRLB) can not be smaller than the sum of the FIM's reciprocal diagonal elements, based on the inequality transformation method in \cite{optimal}, the optimal objective becomes  
\begin{equation}\label{object}
	\mathop{\arg\min}\limits_{\left\{d_i^l\right\},i=1,\cdots,N_l} tr({\rm{CRLB}}) \geqslant \frac{4}{\Phi_{11}+\Phi_{22}}.
\end{equation}
To obtain the simplified result of $\Phi_{11}+\Phi_{22}$, a special situation is considered, where the VUE's trajectory and positions of all BSs are on a line. The lowest trace values of CRLB in other scenarios will be greater than the value in this scenario. In this situation, $\cos{\theta_{i+1,k}^l} = \cos{\theta_{i,k}^l}$ and $\sin{\theta_{i+1,k}^l}=\sin{\theta_{i,k}^l}$. Since the signal sampling distance interval $d_{i+1,k}^l-d_{i,k}^l = 1$ m, and there is a conversion relationship between the mean square gradient variance and the RSS variance, by plugging (\ref{jacobi}) - (\ref{c1}) into (\ref{object}), the smallest tr(CRLB) is a value related to the distances between VUE and BSs, and we have
\begin{equation}\label{result}
 tr({\rm{CRLB}}) \geqslant \frac{{\ln^2{10}}}{\frac{{4\cdot10^4}}{\left(N_l-1\right)^2}{\sum_{k=1}^{K}\frac{{\beta_k^l}^4}{{\rho_k^l}^2}(\frac{1}{d_{N_l,k}^l}-\frac{1}{d_{1,k}^l}\frac{1}{N_l})^2}}.
\vspace{-5mm}
\end{equation}
%\vspace{-5mm} 
\vspace{2mm}
\section*{Appendix B: \textsc{Proof of Proposition 1}}
\label{ProPropo2}
$C_{g^2}(d)$ is defined as the CRLB of the $l$th segment whose endpoints are not the signal extreme points. If the $l$th segment is partitioned by the signal extreme points, i.e., $d_{N_l,k}^l= d_{min,k}^l$ and ${d_{1,k}^{l}}= d_{max,k}^l$, the CRLB of this $l$th segment is expressed as $C_{g^2}(d^{'})$. The ratio between $C_{g^2}(d)$ and $C_{g^2}(d^{'})$ can be given as follows: 
\begin{equation}\label{result2}
\begin{split}
&{\frac{C_{g^2}(d)}{C_{g^2}(d^{'})}}=\frac{{\ln^2{10}}}{\frac{{4\cdot10^4}}{\left(N_l-1\right)^2}{\sum_{k=1}^{K}\frac{{\beta_k^l}^4}{{\rho_k^l}^2}(\frac{1}{d_{N_l,k}^l}-\frac{1}{d_{1,k}^l}\frac{1}{N_l})^2}} \times \\&
\frac{\frac{{4\cdot10^4}}{\left(N_l-1\right)^2}{\sum_{k=1}^{K}\frac{{\beta_k^l}^4}{{\rho_k^l}^2}(\frac{1}{d_{min,k}^l}-\frac{1}{d_{max,k}^l}\frac{1}{N_l})^2}}{{\ln^2{10}}}\\&
=\frac{\sum_{k=1}^{K}\frac{{\beta_k^l}^4}{{\rho_k^l}^2}(\frac{1}{d_{min,k}^l}-\frac{1}{d_{max,k}^l}\frac{1}{N_l})^2}{\sum_{k=1}^{K}\frac{{\beta_k^l}^4}{{\rho_k^l}^2}(\frac{1}{d_{N_l,k}^l}-\frac{1}{d_{1,k}^l}\frac{1}{N_l})^2}.
\end{split}
\end{equation}   
Since the segment corresponding to $C_{g^2}(d)$ is not partitioned by the signal extreme points, its distances of endpoints are not the extreme points as well. Thus, ${d_{N,k}^l} > d_{min,k}^l$ and ${d_{1,k}^l}<d_{max,k}$. Then, ${\frac{C_{g^2}(d)}{C_{g^2}(d^{'})}}>1$, which proves the proposition 1.
\vspace{-3.5mm}
\section*{Appendix C: \textsc{Proof of Proposition 2}}
\label{fusedCRLB}
The mean RSS feature of the $k$th BS within the $l$th segment can be expressed as follows:
\begin{equation}\label{mu}
\mu_k^l = \frac{1}{N_l}\sum_{i=1}^{N_l}P_{i,k}^l=\frac{-10{\beta_k^l}}{N_l}\sum_{i=1}^{N_l}\log{d_{i,k}^l}+P_{0,k}^l+{W_k^l}
\end{equation}
where $d_{i,k}^l$ is the distance between the VUE's $i$th position and the $k$th BS's position. By plugging (\ref{distance}) - (\ref{mid2}) into (\ref{mu}), the derivative of the $k$th BS's mean RSS feature can be expressed by 
\begin{equation}\label{muderiv}
	\vspace{-1mm}
	\frac{\partial {\mu_{k}^l} }{\partial {x_m^l}} = 
	\frac{-20{\beta_k^l}}{(N_l^2)\ln 10}\sum\limits_{i = 1}^{N_l} \frac{i\cos{\theta_{i,k}^l}}{d_{i,k}^l},
	\vspace{1mm}
\end{equation}
\begin{equation}
	\vspace{-1mm}
	\frac{\partial {\mu_{k}^l} }{\partial {y_m^l}} = 
	\frac{-20\beta_k^l}{(N_l^2)\ln 10}\sum\limits_{i = 1}^{N_l} \frac{i\sin{\theta_{i,k}^l}}{d_{i,k}^l}.
	\vspace{-1mm}
\end{equation}
The Jacobian matrice of the all BSs measurement errors evaluated at the true segment $\boldsymbol{c}_m^l = \left[ x_m^l, y_m^l \right]^\top$ is 
\begin{equation}\label{mujacobi}
	\vspace{-1mm}
	{J_{\mu}} = 
	\begin{pmatrix}
		\frac{\partial {\mu_{1}^l}}{\partial {x_m^l}} & \frac{\partial {\mu_{1}^l}}{\partial {y_m^l}} \\
		\vdots & \vdots \\
		\frac{\partial {\mu_{K}^l}}{\partial {x_m^l}}&
		\frac{\partial {\mu_{K}^l}}{\partial {y_m^l}}	
	\end{pmatrix} = \alpha 
	\begin{pmatrix}
		\sum\limits_{i = 1}^{N_l} \frac{i\cos{\theta_{i,1}^l}\beta_1^l}{d_{i,1}^l} & \sum\limits_{i = 1}^{N_l} \frac{i\sin{\theta_{i,k}^l}\beta_1^l}{d_{i,k}^l} \\
		\vdots & \vdots \\
		\sum\limits_{i = 1}^{N_l} \frac{i\cos{\theta_{i,K}^l}\beta_K^l}{d_{i,K}^l}&
		\sum\limits_{i = 1}^{N_l} \frac{i\sin{\theta_{i,K}^l}\beta_K^l}{d_{i,K}^l}	
	\end{pmatrix}
	\vspace{-1mm}
\end{equation}
where $\alpha=\frac{-20}{(N_l^2)\ln 10}$.
The measurement noise covariance matrix of the mean RSS feature is
\begin{equation}\label{munoise}
	\Sigma{\mu} = diag [ {\eta_1^l}^2,\cdots,{\eta_K^l}^2]_{N_l\times N_l}. 
\end{equation}
The Fisher information matrix (FIM) of the mean RSS feature and the mean square gradient feature takes the form
\begin{equation}\label{cmug}
	\Phi_{\overline{g^{_2}},\mu} = J_{\overline{g^{_2}}}^\top \Sigma_{\overline{g^{_2}}}^{-1} J_{\overline{g^{_2}}}+
	J_{\mu}^\top \Sigma_{\mu}^{-1} J_{\mu}.
\end{equation}
The corresponding CRLB of the joint features can be obtained by
\begin{equation}
\vspace{-1mm}
	{\rm{CRLB}}_{\overline{g^{_2}},\mu} = \Phi_{\overline{g^{_2}},\mu}^{-1},
\vspace{-1mm}
\end{equation}
In (\ref{cmug}), the FIM of the mean RSS feature can be expressed as
\begin{equation}\label{phimu}
\vspace{-1mm}
	\Phi_{\mu} = J_{\mu}^\top \Sigma_{\mu}^{-1} J_{\mu},
\vspace{-1mm}
\end{equation}
And the FIM of mean square gradient feature in (\ref{cmug}) can be expressed as
\begin{equation}\label{phig2}
\vspace{-1mm}
	\Phi_{\overline{g^{_2}}} = J_{\overline{g^{_2}}}^\top \Sigma_{\overline{g^{_2}}}^{-1} J_{\overline{g^{_2}}},
\vspace{-1mm}
\end{equation}
Based on the inequality transformation method to obtain the lowest CRLB,  ${\rm{CRLB}}_{\overline{g^{_2}},\mu}$ can be obtained as follows:
\begin{equation}\label{CRLBmug2}
	tr({\rm{CRLB}}_{\overline{g^{_2}},\mu}) \geqslant \frac{4}{\Phi_{\mu,11}+\Phi_{\mu,22}+\Phi_{\overline{g^{_2}},11}+\Phi_{\overline{g^{_2}},22}},
\end{equation}
Based on only the mean square gradient feature, the CRLB of the segment positioning takes the form 
\begin{equation}\label{CRLBg2}
	tr({\rm{CRLB}}_{\overline{g^{_2}}}) \geqslant \frac{4}{\Phi_{\overline{g^{_2}},11}+\Phi_{\overline{g^{_2}},22}},
\end{equation}
Since ${\Phi_{\mu,11}+\Phi_{\mu,22}+\Phi_{\overline{g^{_2}},11}+\Phi_{\overline{g^{_2}},22}}>{\Phi_{\overline{g^{_2}},11}+\Phi_{\overline{g^{_2}},22}}$, thus $tr({\rm{CRLB}}_{\overline{g^{_2}},\mu}) < tr({\rm{CRLB}}_{\overline{g^{_2}}})$. The proof of Proposition 2 is completed. 
\vspace{-4mm}
\section*{Appendix D: \textsc{Proof of Proposition 3}}
\label{CRLBBSK}
Suppose the $K$th BS's signals of the $l$th segment change within a specified threshold, thus the distances of positions in this segment equivalently satisfy $d_{1,K}^l \approx d_{2,K}^l \cdots \approx d_{N_l,K}^l$. By plugging (\ref{deriv1}) and (\ref{deriv2}) into (\ref{phig2}), the trace element of the $\Phi_{\overline{g^{_2}}}$ can be expressed as follows:
\begin{equation}\label{phig2trace1}
	\begin{split}
		\vspace{-1mm}
		&\Phi_{\overline{g^{_2}},11}=\frac{16\cdot10^4}{\left(N_l-1\right)^2\ln^2{10}}{\sum\limits_{k = 1}^{K} \frac{{\beta_k^l}^4}{{\rho_k^l}^2}} [\sum\limits_{i = 1}^{N_l-1}{\left( \log {d_{i+1,k}^l}-\log {d_{i,k}^l} \right)} \\&		 
		\left( \frac{\cos{\theta_{i+1,k}^l}}{d_{i+1,k}^l}
		\frac{(i+1)}{N_l}-\frac{\cos{\theta_{i,k}^l}}{d_{i,k}^l}
		\frac{i}{N_l} \right)]^2,				
		\vspace{-2mm}
	\end{split}
\end{equation}
\begin{equation}\label{phig2trace2}
	\begin{split}
		\vspace{-1mm}
		&\Phi_{\overline{g^{_2}},22}\frac{16\cdot10^4}{\left(N_l-1\right)^2\ln^2{10}}{\sum\limits_{k = 1}^{K} \frac{{\beta_k^l}^4}{{\rho_k^l}^2}} [\sum\limits_{i = 1}^{N_l-1}{\left( \log {d_{i+1,k}^l}-\log {d_{i,k}^l} \right)} \\&		 
		\left( \frac{\sin{\theta_{i+1,k}^l}}{d_{i+1,k}^l}
		\frac{(i+1)}{N_l}-\frac{\sin{\theta_{i,k}^l}}{d_{i,k}^l}
		\frac{i}{N_l} \right)]^2.				
		\vspace{-4mm}
	\end{split}
\vspace{-3mm}
\end{equation}
\vspace{-0.5mm}
Since the $d_{1,K}^l \approx d_{2,K}^l \cdots \approx d_{N_l,K}^l$, thus $\sum\limits_{i = 1}^{N_l-1}(\log {d_{i+1,K}^l}-\log {d_{i,K}^l})=0$. By plugging  (\ref{phig2trace1}) and (\ref{phig2trace2}) into (\ref{CRLBg2}), due to $\sum\limits_{i = 1}^{N_l-1}(\log {d_{i+1,K}^l}-\log {d_{i,K}^l})=0$, thus the ${\rm{CRLB}}(C_K)$ = ${\rm{CRLB}}(C_{K-1})$. The proof of Proposition 3 is completed.    

\footnotesize  	
\bibliography{mybibfile}
\bibliographystyle{ieeetr}

\clearpage
\normalsize

\end{document}